\documentclass[submission,letterpaper, Phys]{SciPost}

\pdfoutput=1
\usepackage{amsmath,amssymb,mathtools,xspace}
\usepackage{booktabs,multirow,graphicx,subcaption,tabularx,slashed}
\usepackage{hyperref}
\usepackage{color,xcolor}
\usepackage[normalem]{ulem}
\usepackage{enumitem}

\makeatletter
\@ifundefined{pdfoutput}{}{\DeclareGraphicsRule{*}{mps}{*}{}}
\makeatother

\newcommand\one{\leavevmode\hbox{\small1\normalsize\kern-.33em1}}

\newcommand{\mdm}{m_\chi}

\def\slashchar#1{\setbox0=\hbox{$#1$}           
   \dimen0=\wd0                                 
   \setbox1=\hbox{/} \dimen1=\wd1               
   \ifdim\dimen0>\dimen1                        
      \rlap{\hbox to \dimen0{\hfil/\hfil}}      
      #1                                        
   \else                                        
      \rlap{\hbox to \dimen1{\hfil$#1$\hfil}}   
      /                                         
   \fi}

\newcommand{\eg}{\textsl{e.g.}\;}
\newcommand{\ie}{\textsl{i.e.}\;}

\begin{document}

\begin{center}{\Large \textbf{
Hadronic Footprint of GeV-Mass Dark Matter 
}}\end{center}

\begin{center}
Tilman Plehn\textsuperscript{1},
Peter Reimitz\textsuperscript{1}, and
Peter Richardson\textsuperscript{2,3}
\end{center}

\begin{center}
{\bf 1} Institut f\"ur Theoretische Physik, Universit\"at Heidelberg, Germany\\
{\bf 2} Theoretical Physics Department, CERN, Geneva, Switzerland\\
{\bf 3} Institute for Particle Physics Phenomenology, Durham University, UK\\
reimitz@thphys.uni-heidelberg.de
\end{center}

\begin{center}
\today
\end{center}


\section*{Abstract}
{\bf GeV-scale dark matter is an increasingly attractive target for
  direct detection, indirect detection, and collider searches. Its
  annihilation into hadronic final states produces a challenging zoo
  of light hadronic resonances. We update Herwig7 to study 
  the photon and positron spectra from annihilation through a vector
  mediator. It covers dark matter masses between 250 MeV and 5 GeV
  and includes an error estimate.}

\vspace{10pt}
\noindent\rule{\textwidth}{1pt}
\tableofcontents\thispagestyle{fancy}
\noindent\rule{\textwidth}{1pt}
\vspace{10pt}

\newpage
\section{Introduction}
\label{sec:intro}

The fundamental nature of dark matter is the biggest particle physics
question of our time. It follows directly from the success of quantum
field theory in describing the properties of elementary particles, as
well as the standard cosmology after the Big Bang. A problem is that
the term particle dark matter is very loosely defined, covering very
light particles essentially giving a background wave function all the
way to primordial black holes. Theory motivations are widely used to
support certain mass ranges, but given the generally modest success of
models for physics beyond the Standard Model they should be taken with
a truck load of salt~\cite{Lin:2019uvt}.

The defining feature of dark matter particles --- closely related
to the one actual measurement of the relic density~\cite{Ade:2015xua}
--- is the dark matter production mechanism. It needs to explain the
observed relic density in agreement with the largely known thermal
history of the universe. For wide classes of dark matter models this
implies an interaction with SM particles beyond an obviously existing
gravitational interaction. Dark matter masses around the weak scale
and down to the GeV scale can be produced thermally through freeze-out
or freeze-in, or through a relation with the baryon asymmetry in the
universe. These mechanisms require more or less strong couplings to
the SM matter particles, \ie to leptons or quarks.  In this mass range
there exists a wealth of relatively model-independent direct detection
constraints~\cite{Arcadi:2017kky}, and their extension to lighter dark
matter particles at and below the GeV-scale is one of the most
interesting experimental
directions~\cite{Agnes:2018ves,Aprile:2016wwo,Essig:2017kqs}. For such
GeV-scale dark matter especially the couplings to quarks are not well
constrained.

Indirect searches for dark matter are another way to directly probe
the properties of the dark matter in the universe. A leading
signature are photons produced in the annihilation of dark matter in
dense regions of the
sky~\cite{Gunn:1978gr,Stecker:1978du,Zeldovich:1980st,Ellis:1988qp}. 
These photons can constrain dark matter interactions with the Standard
Model on the lepton and the hadron side. The reason is that they can
be produced directly and through radiation from any charged
annihilation product. Just as for positrons, we know the photon
spectrum from many particle physics experiments over the recent
decades. They are available for instance through the \textsc{Pppc4dmid}
tool~\cite{Cirelli:2010xx} based on
\textsc{Pythia}~\cite{Sjostrand:2007gs}.  Standard tools like
\textsc{micrOMEGAs}~\cite{Belanger:2001fz,Belanger:2018mqt},
\textsc{MadDM}~\cite{Backovic:2013dpa,Ambrogi:2018jqj}, or
\textsc{DarkSusy}~\cite{Gondolo:2004sc,Bringmann:2018lay} include
similar spectra based on multi-purpose Monte Carlo generators. A major
technical problem with dark matter annihilation into hadrons is that
its description is not available through \textsc{Pythia} once the dark
matter masses drop below around 5~GeV.  The only exception is the
recent \textsc{Hazma}~\cite{Coogan:2019qpu} tool for dark matter
masses below 250~MeV~\cite{Sabti:2019mhn}.  This leaves dark matter
annihilation to the leading hadronic final states for masses between
250~MeV and 5~GeV essentially uncovered.\bigskip

Technically, GeV-scale dark matter annihilation through light scalar
or light vector mediators is very different. If we assume that a new
scalar couples to SM particles with Yukawa couplings roughly reflecting
the SM mass hierarchy, increasingly weak GeV-scale dark matter will
annihilate into charm quarks and tau leptons, followed by muons and
eventually pions and electrons. From a hadronic physics point of view
the more interesting scenario are vector mediators, where the SM
interactions are generation-universal. In that case we will observe a
wealth of hadronic annihilation channels below the $b\bar{b}$
threshold. These annihilation channels will have distinct photon and lepton 
spectra, which we will focus on in this study.

Finally, the proper description of dark matter annihilation to
hadronic final state is plagued by large uncertainties, as for
instance pointed out for \textsc{Pppc4dmid}~\cite{Cirelli:2010xx} in
relation to \textsc{Pythia}~\cite{Sjostrand:2007gs} and
\textsc{Herwig}~\cite{Corcella:2000bw}.  Consequently, dedicated
comparisons between \textsc{Pythia} and \textsc{Herwig} have been
published for dark matter annihilation to tau leptons, bottom and top
quarks, and weak bosons~\cite{Cembranos:2013cfa}.  More recently, this
comparison has been updated~\cite{Niblaeus:2019ldk} to the most recent
versions of \textsc{Pythia}8~\cite{Sjostrand:2007gs,Sjostrand:2014zea}
and \textsc{Herwig}7~\cite{Bellm:2015jjp,Bellm:2017bvx}.  A detailed
analysis of the the \textsc{Pythia} predictions can be found in
Ref.~\cite{Amoroso:2018qga}. All of these studies target relatively
heavy dark matter annihilation, in line with the common weakly
interacting massive particle~(WIMP) hypothesis.

In this paper we provide the first proper description of photon and
lepton spectra from GeV dark matter annihilating into hadronic final
states based on \textsc{Herwig} with an updated fit to
electron--positron data, including several new final states. They
become relevant when we reduce the dark matter scattering energy below
the \textsc{Pythia} limit. We update the fit to electron--positron
data as the input to the \textsc{Herwig} description and add the
necessary new hadronic final states with up to four
hadrons. Especially for the photon spectrum we observe a complete
change in the shape of the spectrum when we reduce the dark matter
mass, starting from typical hadron decay chains to continuum
multi-pion production. In addition we provide a first estimate of the
impact of the input-data fit uncertainties on the output
spectra.\bigskip

The paper is structured the following way: after introducing our toy
model in Sec.~\ref{sec:model} we review the established
implementations in Sec.~\ref{sec:status}. We show how their
reliability starts to fade once we go below dark matter masses of
5~GeV and the tools start to extrapolate beyond their common
\textsc{Pythia} input. In Sec.~\ref{sec:herwig} we show the results
from our new \textsc{Herwig}-based implementation. We focus on shape
changes in the photon and lepton spectra when we reduce the dark
matter mass towards into the continuum-pion regime. We also show the
error bands on the photon and positron spectra from the fit
uncertainties to the electron-positron input data.  In the Appendix we
provide all details about our new fit, the underlying parametrizations,
the best-fit points, and the error bands.

\section{Toy model}
\label{sec:model}

The standard interpretation framework for weak-scale dark matter is
thermal freeze-out production or the WIMP paradigm.  Embedding GeV-scale
dark matter searches in a global analysis~\cite{Leane:2018kjk}
provides an excellent illustration of the many cosmological
constraints and their model dependence. Above masses around 10~GeV,
FERMI constrains these scenarios using photons in dwarf spheroidal
galaxies~\cite{Ackermann:2015zua,Fermi-LAT:2016uux}, while AMS covers
 leptonic final states~\cite{Aguilar:2014mma,Accardo:2014lma}. The
\textsc{e-ASTROGAM} program~\cite{DeAngelis:2017gra}, for example, is proposing searches for gamma-rays in
the MeV-GeV region. In
addition, precision measurements of the Cosmic Microwave Background~(CMB)~\cite{Ade:2015xua} are
sensitive to the total ionizing energy either directly (electrons and
muons) or indirectly. Finally, Big-Bang Nucleosynthesis~(BBN) does not allow WIMPs below around
10~MeV~\cite{Nollett:2013pwa,Nollett:2014lwa}. The main difference
between these different analyses is that some rely on
assumptions on the thermal history of dark matter. 

For DM masses in the range $\mdm = 0.1~...~7$~GeV the CMB provides the
leading constraint on GeV-scale dark matter, where asymmetric dark
matter as an alternative production model leads to weaker CMB
constraints if the dark matter is sufficiently
asymmetric~\cite{Lin:2011gj}. For an anti-DM to DM ratio of less than
$\sim 2\times 10^{-6} (10^{-1})$ for DM masses $m_\chi = 1~\text{MeV}
(10~\text{GeV})$, CMB constraints can be evaded as seen in Fig. 1
of~\cite{Lin:2011gj}. Not yet being fully asymmetric, one still gets
indirect detection signals. Other
modifications at least weakening the CMB constraints for thermal
production are softer spectra from annihilation modes beyond $2 \to
2$ kinematics~\cite{Elor:2015bho,Elor:2015tva}, including a dominant
$2 \to 3$ bremsstrahlung
process~\cite{Bergstrom:1989jr,Flores:1989ru,Baltz:2002we,Bringmann:2007nk,Bergstrom:2008gr,Barger:2009xe,Bell:2010ei,Bell:2011if,Bell:2011eu,Ciafaloni:2011sa,Garny:2011cj,Bell:2012dk,DeSimone:2013gj,Ciafaloni:2013hya,Bell:2017irk,Bringmann:2017sko}. However,
the necessary annihilation rate is typically too small to lead to
observed relic density.\bigskip

To define a toy model for our hadronization study we assume that the
observed dark matter density is somehow produced through thermal
freeze-out, but with a light vector mediator. We assume the dark matter
candidate to be a Majorana fermion $\chi$, but our results apply the
same way to asymmetric dark matter where the dark matter fermion has
to be different from its anti-particle. Since our study is based on
$e^+e^-$- data, we have to focus on vector mediator models. A simple mediator choice
starts from an additional $U(1)$ gauge symmetry, where we gauge one of
the accidental global symmetries related to baryon and lepton
number~\cite{Chang:2000xy,Chen:2006hn,Salvioni:2009jp,Lee:2010hf,Araki:2012ip}. For
our purpose of testing dark matter annihilation into light-flavor jets
with a limited number of photons from leptonic channels the most
attractive combination is $B - 3 L_\mu$~\cite{Heeck:2018nzc}. This
gives us the annihilation channel
\begin{align}
\chi \chi \to Z' \to \text{Standard Model} \; .
\end{align}
To avoid strong biases from an underlying model we also show results
for $Z'$ couplings similar to the Standard Model
case for low energies.  For
consistent field theory models the annihilation to SM quarks will
always occur at the loop level, even if they are suppressed at tree
level~\cite{DEramo:2017zqw}. As our benchmark model we therefore
assume an approximately on-shell annihilation
\begin{align}
\chi \chi \to Z' \to q \bar{q}
\qquad \text{with} \quad 
m_{Z'} \approx 2 m_\chi \; .
\label{eq:model}
\end{align}
The coupling strength of the DM to the mediator can be chosen
arbitrarily, because we are only interested in the form of the energy
spectra from the hadronic final states. 
For light dark matter masses the relevant
quarks are $u,d,s,c$, possibly the bottom quark. The charm quark plays
a special role, because threshold region is poorly
understood. Examples for distinct photon spectra from annihilations to
$c$ and $b$ quarks are, for example, described
in~\cite{Bringmann:2016axu}. All we can do is
rely on the spectra included in \textsc{Pythia} or \textsc{Herwig},
with little improvement on the modelling side. \bigskip

For the three lightest
quarks there exists a wealth of measurements which we can use to
constrain dark matter annihilation into hadrons. We decompose a quark
DM current $J_\text{DM}^\mu=\sum_{q=u,d,s} a_q \bar{q}\gamma^\mu q$
into isospin components and a separate $s\bar{s}$ contribution,
\begin{align}
J_\text{DM}^\mu = \frac{1}{\sqrt{2}}\left(
(a_u-a_d)J^{I=1,3,\mu}+(a_u+a_d)J^{I=0,\mu} \right)+a_s J^{s,\mu}
\label{eq:DMcurrent}
\end{align}
where $a_q$ are the couplings of the light vector mediator to the
light quarks, $q=u,d,s$.
The mediator couplings to quarks are fixed
to $a_q=1/3$ for any anomaly-free $B-L$ model.  Depending on the
mediator coupling structure to quarks, one or the other isospin
current might vanish. As a consequence, some resonance contributions
to the channels might vanish, or even more drastically, pure isospin $I=0$ channels, for example
\begin{align}
\chi \chi \to \omega\pi\pi, \eta\omega, \ldots \, .
\label{eq:I0}
\end{align}
or pure $I=1$ channels such as
\begin{align}
\chi \chi \to \pi\pi, 4\pi, \eta\pi\pi, \omega\pi, \phi\pi, \eta'\pi\pi, \ldots
\label{eq:I1}
\end{align}
are absent. We also choose to include the isospin breaking
contribution from $\omega\to\pi^+\pi^-$ in the $I=1$ current for
simplicity.  The general matrix element for DM annihilation can be
written in the form
\begin{align}
\mathcal{M}=a_\text{DM}\bar{v}(p_1)\gamma^\nu
u(p_2) d_{\nu\mu}^\text{DM}\langle X|J_\text{DM}^\mu|0\rangle
\end{align}
with the DM-mediator coupling $a_\text{DM}$ and the vector
mediator propagator
$d_{\nu\mu}^\text{DM}$. 

In our toy model we always assume $m_{Z'} = 2 \mdm$, but given the
non-relativistic nature of DM annihilation the mediator mass should
only have negligible impact on our spectra. Since the mass of the
mediator determines the width of the mediator, we calculate the width
in the hadronic resonance region within \textsc{Herwig} through its
decays to all kinematically allowed hadronic final states listed in
Tab.~\ref{tab:channels} of the Appendix.

\section{Established tools}
\label{sec:status}

\begin{table}[b!]
\setlength{\tabcolsep}{12pt}
\centering
\begin{tabular}{l|lrl}
\toprule
Tool & Back-end & $\mdm^\text{min}$ & DM models \\ \midrule
\textsc{Pppc4dmid}    & \textsc{Pythia}8.135 tables &  5 GeV & generic DM\\[2mm]
\textsc{micrOMEGAs} & \textsc{Pythia}6.4 tables & $\sim 2$~GeV & UFO model\\[2mm]
\multirow{2}{*}{\textsc{MadDM}} & \textsc{Pppc4dmid}   & 5 GeV & \multirow{2}{*}{UFO model} \\
                                & \textsc{Pythia}8.2 direct & $\sim 2$~GeV & \\[2mm]
\textsc{DarkSusy} & \textsc{Pythia}6.426 tables & $\sim 3$~GeV & generic DM, SUSY\\[2mm]
\bottomrule
\end{tabular} 
\label{tab:comparison}
\caption{Comparison of publicly available tools to generate spectra
  from DM annihilation.}
\end{table}

Different public tools generate energy
spectra for different DM annihilation channels to SM particles. They
are limited in DM masses by their approach and by their back-end, but
are mediator-independent. We
summarize
\begin{itemize}
\item \textsc{Pppc4dmid}~\cite{Cirelli:2010xx} provides tabulated energy
  spectra for indirect detection. The $e^\pm$, $\bar{p}$, $\bar{d}$,
  $\gamma$, and $\nu_{e,\mu,\tau}$ fluxes are generated with
  \textsc{Pythia}8.135~\cite{Sjostrand:2007gs} down to $\mdm =
  5$~GeV. We use the provided interpolation routine to extrapolate the
  results to $\mdm = 2$~GeV.
\item \textsc{micrOMEGAs}~\cite{Belanger:2001fz,Belanger:2018mqt} uses
  tabulated \textsc{Pythia} spectra for
  $\gamma,e^+,\bar{p},\nu_{e,\mu,\tau}$ and extrapolates down to $\mdm
  = 2$~GeV. In the manual of version \textsc{micrOMEGAs}2.0 it is
  mentioned that the strategy for calculating spectra is analogous to
  that of \textsc{DarkSusy} and that spectra extrapolated to masses
  below 2~GeV should be taken with care.
\item \textsc{MadDM}~\cite{Backovic:2013dpa,Ambrogi:2018jqj} provides
  two ways of calculating the energy spectra both based on
  \textsc{Pythia}~\cite{Sjostrand:2014zea}. The `fast' calculation is
  based on the numerical tables provided by \textsc{Pppc4dmid}. In the
  `precise' mode, events are generated with \textsc{MadGraph} and then
  passed to \textsc{Pythia} for showering and hadronization. In this
  mode it is possible to calculate the fluxes of any final states
  based on the UFO model implementation.
\item \textsc{DarkSusy}~\cite{Gondolo:2004sc,Bringmann:2018lay}
  provides tables down to 3~GeV for energy spectra of two-particle SM
  final states based on
  \textsc{Pythia}6.426~\cite{Sjostrand:2006za}. The tool can
  interpolate and extrapolate the $\gamma,
  e^+,\bar{p},\bar{d},\pi_0,\nu_{e,\mu,\tau},\mu$ fluxes for all quark
  final states. In addition it includes annihilation to $\mu \mu$,
  $\tau \tau$, gluons, and weak bosons. Dark matter annihilation into
  $e^+e^-$ pairs appears to not be included.
\item \textsc{Hazma}~\cite{Coogan:2019qpu} is a Python toolkit to
  produce energy spectra in the sub-GeV range. It is based on leading
  order chiral perturbation theory and is valid in the non-resonance
  region below $\mdm = 250$~MeV.
\end{itemize}
From this list it is clear that for dark matter masses in the GeV
range all public tools are based on \textsc{Pythia}, one way or
another. Multi-purpose Monte Carlo tools, such as, \textsc{Pythia} or
\textsc{Herwig} can calculate the energy spectra for many hard scattering
processes, followed by hadronization or fragmentation and hadron
decays. In the range we are interested in these spectra are usually
extracted from data, as discussed in the Appendix.

\begin{figure}[t]
\includegraphics[width=1.0\textwidth]{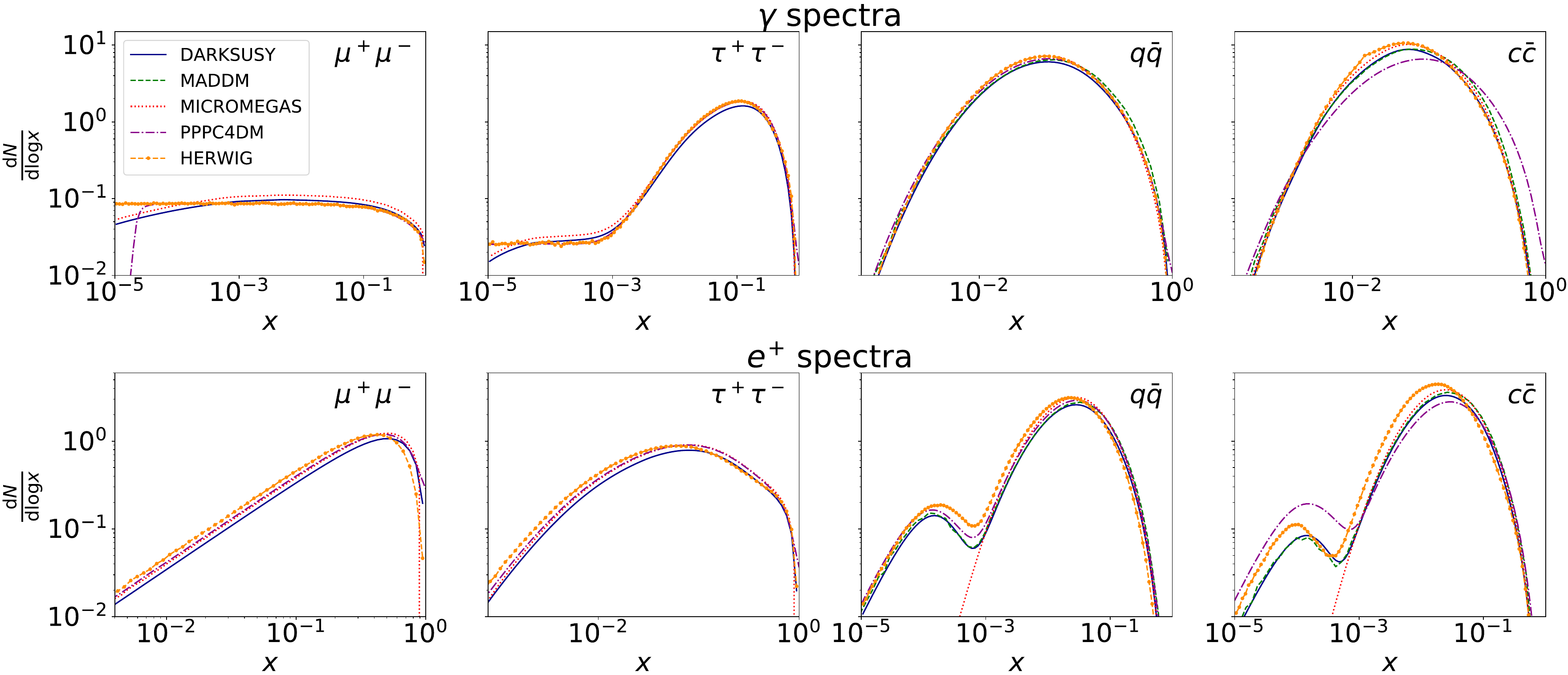}
\caption{Photon and positron spectra $dN /d \log x$ with $x =
  E_\text{kin}/\mdm$ for $\mdm=5$~GeV from the different hard
  annihilation channels. We show results from \textsc{DarkSusy},
  \textsc{MadDM}, \textsc{micrOMEGAs}, \textsc{Pppc4dmid}, and
  \textsc{Herwig}.}
\label{fig:5gev}
\end{figure}

\begin{figure}[b!]
\includegraphics[width=1.0\textwidth]{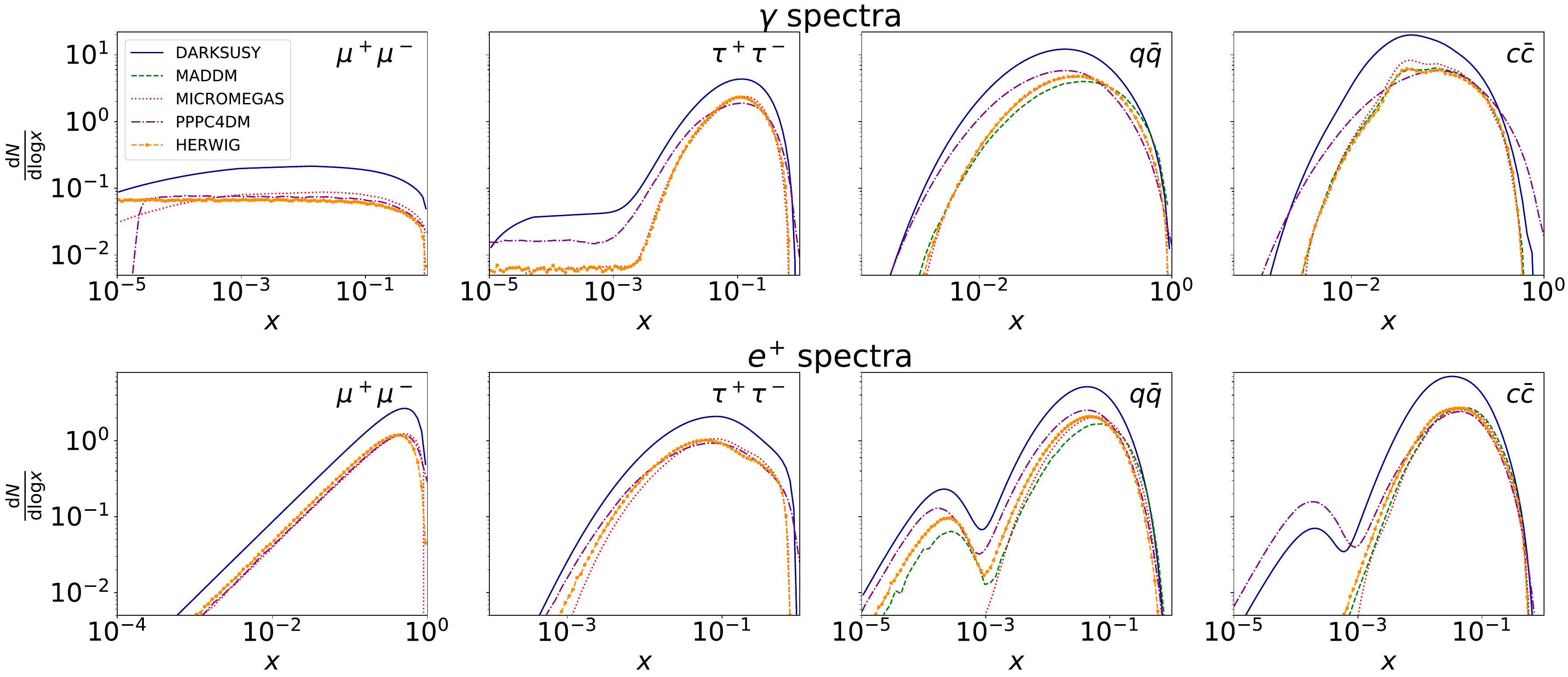}
\caption{Photon and positron spectra $dN /d \log x$ with $x =
  E_\text{kin}/\mdm$ for $\mdm=2$~GeV from the different hard
  annihilation channels. We show results from \textsc{DarkSusy},
  \textsc{MadDM}, \textsc{micrOMEGAs}, \textsc{Pppc4dmid}, and
  \textsc{Herwig}.}
\label{fig:2gev}
\end{figure}

The advantage of the Monte Carlo tools is that we can extract the
cosmologically most relevant photon, lepton, and anti-proton spectra
for each hard dark matter annihilation process. We assume an
annihilation process of the kind given in Eq.\eqref{eq:model}, but
allow for any kinematically allowed SM final state. For the numerical
results we rely on spectra from the processes $e ^+ e^- \to$~SM pairs,
at a given energy $m_{ee} = 2 \mdm$. In Fig.~\ref{fig:5gev} we compare
the corresponding \textsc{Pythia}-like spectra from the standard tools
discussed above. We show the photon and positron spectra from DM
annihilation into muon, tau, and light-quark ($u,d,s$) pairs and
compare them to the standard \textsc{Herwig} output for an alternative
description.  Starting with the left panels of Fig.~\ref{fig:5gev} we
see a flat photon spectrum from soft-enhanced radiation and a
triangular positron spectrum from the $\mu^+$-decay with a
three-particle final state. For taus the hadronic decays produce
neutral and charged pions, where for instance the decay $\pi^0 \to
\gamma \gamma$ dominates down to $x \approx 10^{-3}$. Below that we
again find the flat photon spectrum from soft emission. The dominant
contribution to the position spectrum is the hadronic decay chain
$\tau^+ \to \pi^+ \to \mu^+ \to e^+$, with a sub-dominant contribution
from the leptonic $\beta$-decay $\tau^+ \to e^+$. Next, light-flavor
quarks $u,d,s$ form a range of hadrons which then decay to $\pi^0 \to
2 \gamma$. The positron spectrum from these light quarks includes a soft
neutron $\beta$-decay, which gives rise to the secondary maximum
around $x \approx 10^{-4}$. The neutron decay is not included in our
default version of \textsc{micrOMEGAs}, but can be easily
added. Finally, moving to DM annihilation into charms we see that the
photon and positron spectra are the same as for the light quarks.

In Fig.~\ref{fig:2gev} we show the same spectra, but for a slightly
lower dark matter mass of 2~GeV. This value is slightly beyond where
\textsc{Pythia} output can be used in a straightforward
manner. Essentially all radiation and decay patterns remain the same
as for 5~GeV, but the different curves start moving apart. This is an
effect of individual extrapolations from the \textsc{Pythia}
output. The only interesting feature appears in the annihilation $\chi
\chi \to c \bar{c}$. Here the extrapolated results from
\textsc{Pppc4dmid} and \textsc{DarkSusy} still include a secondary peak
corresponding to the neutron decay in the light quark
channel. However, the lightest charm baryon is $\Lambda_c$ has a mass
of 2.29~GeV, so at $\mdm = 2$~GeV it cannot be produced on-shell. What
we see is likely an over-estimate of off-shell effects or an
extrapolation error from the 5~GeV case, which illustrates the danger
of ignoring the explicit warning not to use for instance
\textsc{Pppc4dmid} or \textsc{DarkSusy} below their recommended mass
ranges. For \textsc{micrOMEGAs} the spectrum is significantly softer
than from the dedicated \textsc{MadDM} call to \textsc{Pythia} and
from \textsc{Herwig}.

Altogether we find that for $\mdm = 5$~GeV there is a completely
consistent picture, where the \textsc{Pythia}-based results are in
excellent agreement with \textsc{Herwig}. Going to $\mdm = 2$~GeV
leads to an increased variation between the different tools and
illustrates why we might not want to use the standard tools outside
their recommended mass ranges.

\section{Herwig4DM spectra}
\label{sec:herwig}

To extend the range of valid simulations of dark matter annihilation
to quarks we start with the standard
\textsc{Herwig}7~\cite{Bahr:2008pv,Bellm:2015jjp,Bellm:2017bvx} implementation. We
then add a set of additional final states and update some other
spectra, as discussed in the Appendix. This allows us to cover DM
masses down to twice the pion mass for vector mediator models. Below the threshold $m_{Z'} = 2
m_\pi \approx 250$~MeV the annihilation to hadrons will be suppressed
and the annihilation to electrons and photons will dominate.
In Fig.~\ref{fig:herwig} we show the photon and positron
spectra from the annihilation process
\begin{align}
\chi \chi \to Z' \to q \bar{q}
\qquad \text{with} \qquad 
q = u,d,s,c
\label{eq:model2}
\end{align}
for decreasing DM masses from $\mdm = 2$~GeV to 250~MeV.

\subsubsection*{Spectra}

\begin{figure}[t]
\centering
\includegraphics[width=0.49\textwidth]{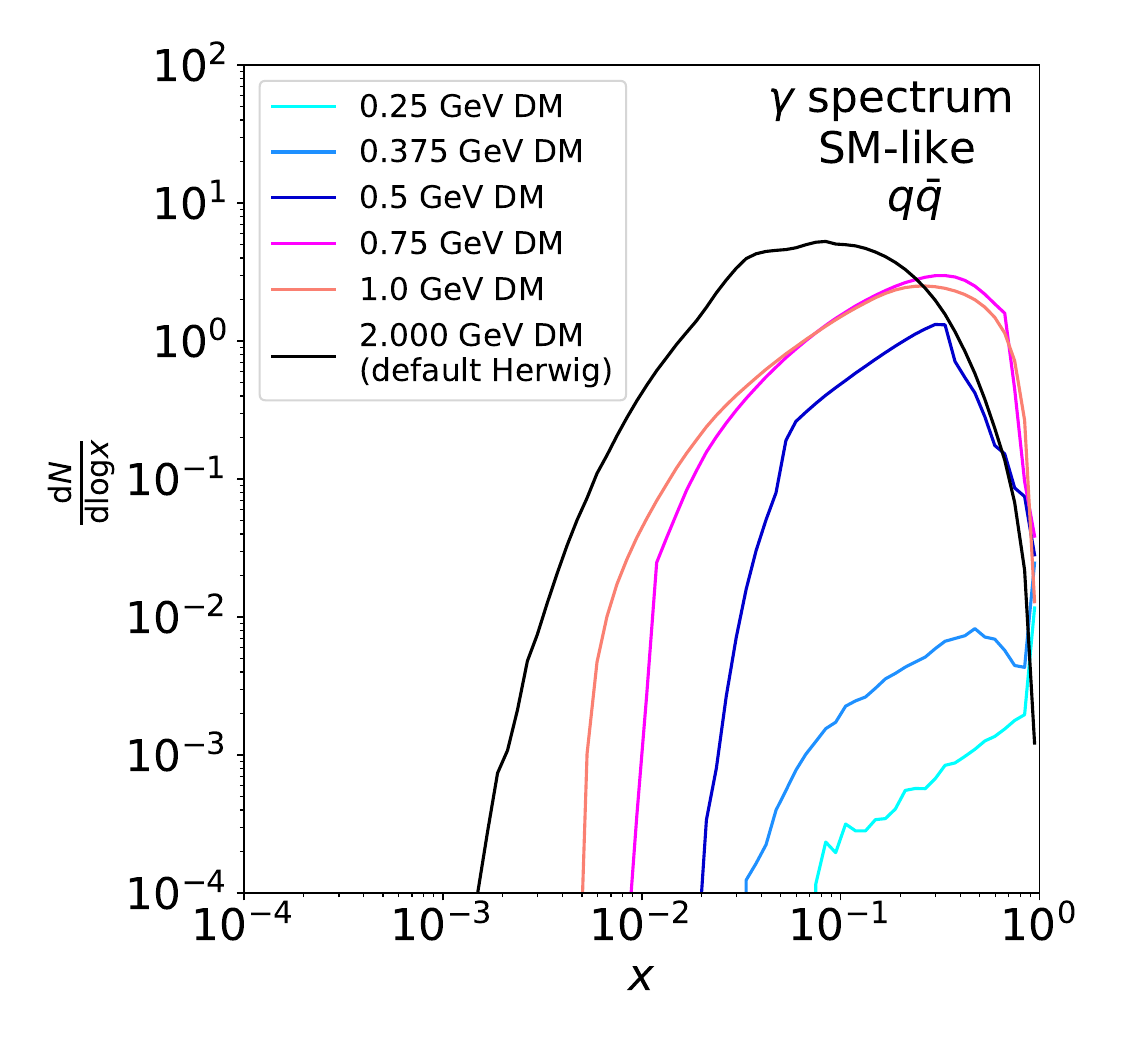}
\includegraphics[width=0.49\textwidth]{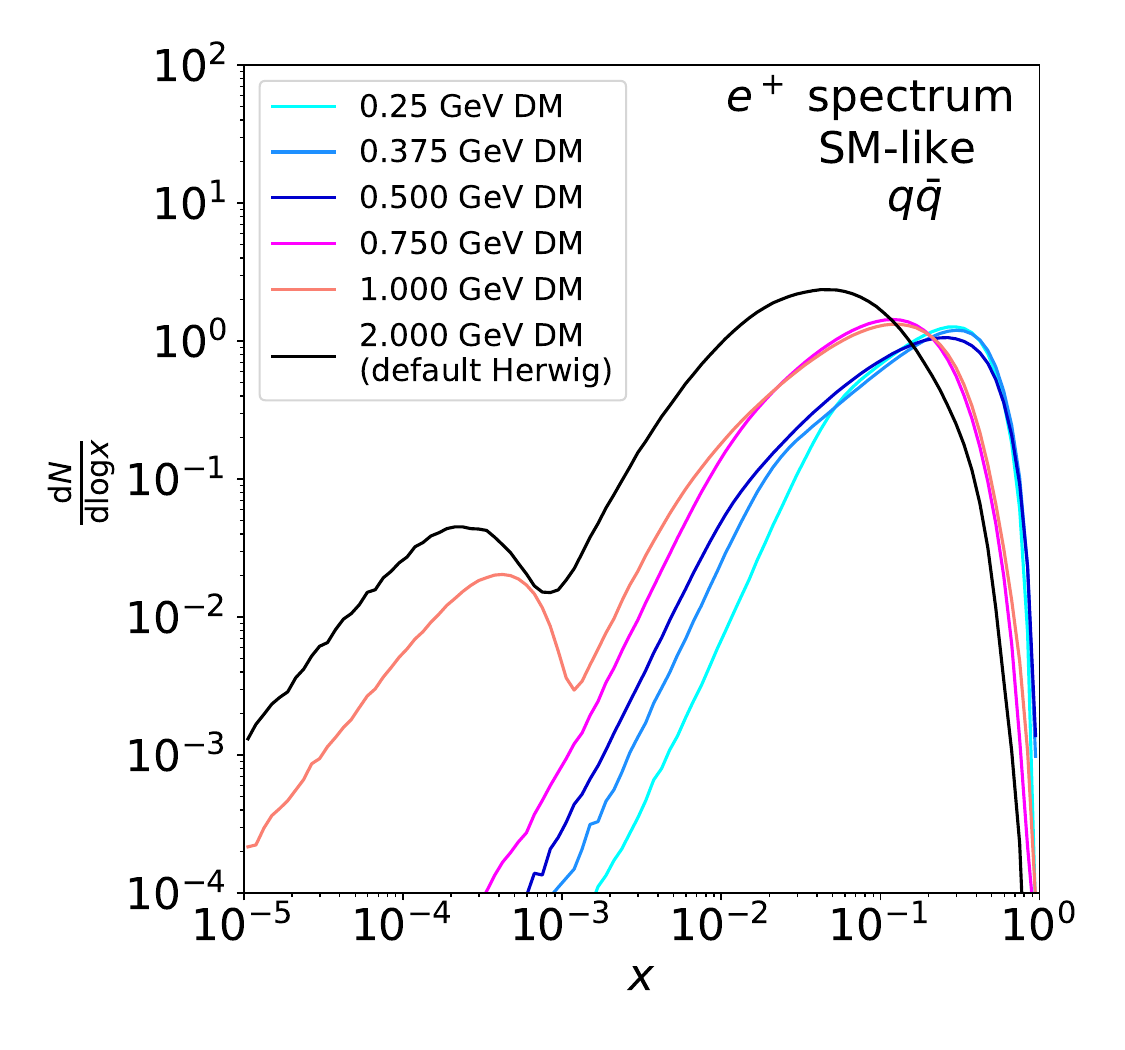} \\
\includegraphics[width=0.49\textwidth]{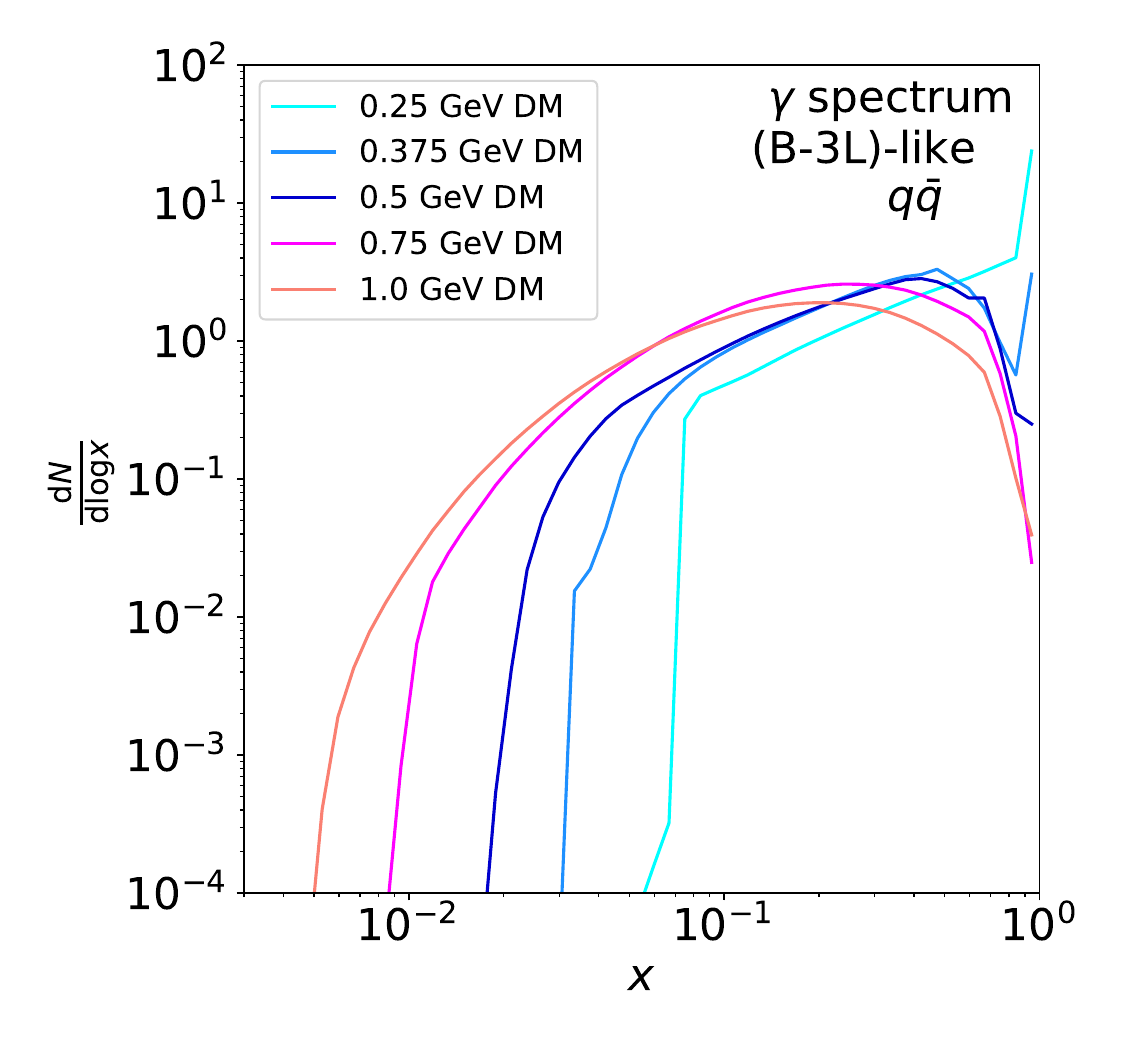}
\includegraphics[width=0.49\textwidth]{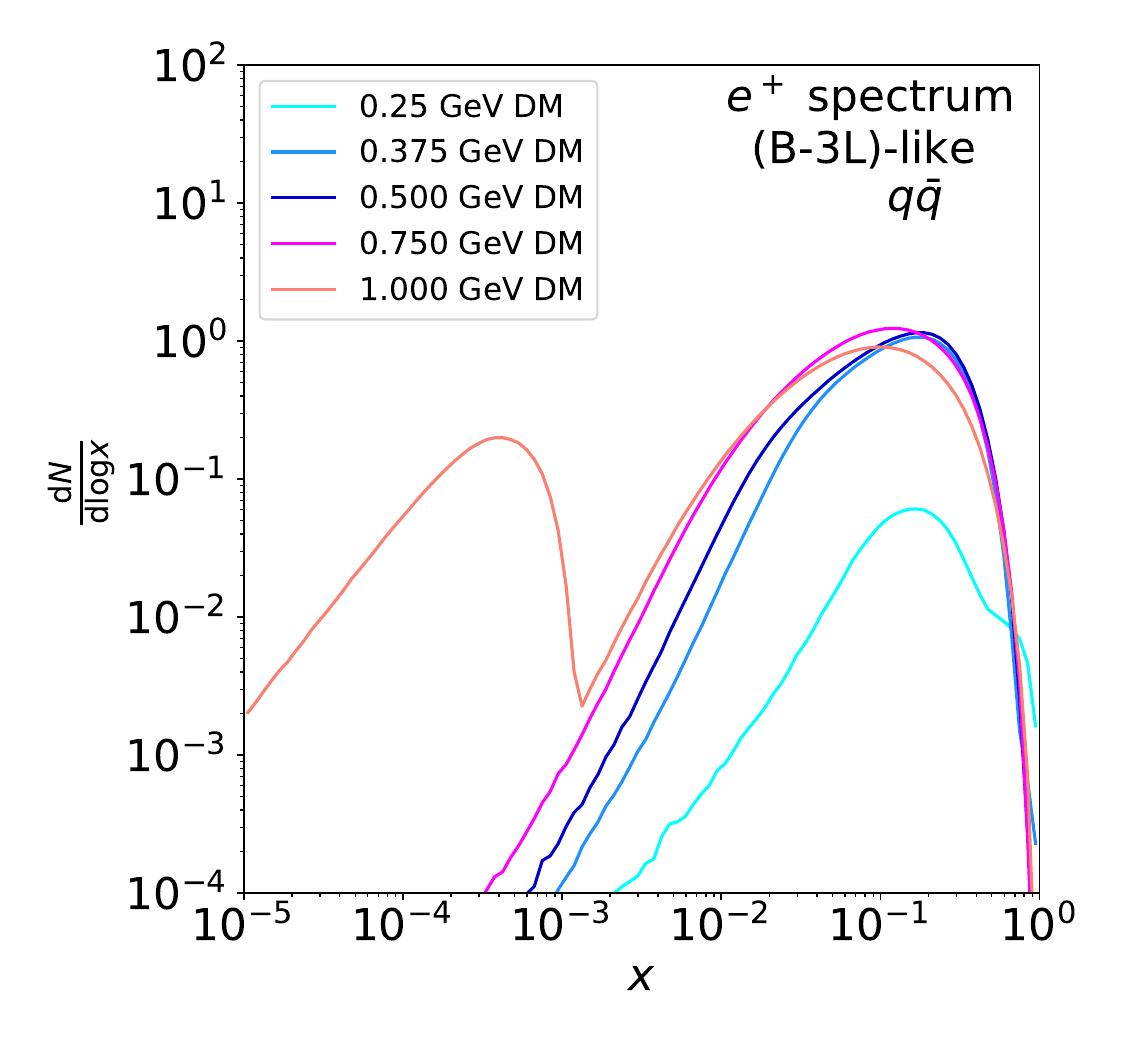}
\caption{Photon and positron spectra $dN /d \log x$ with $x =
  E_\text{kin}/\mdm$ for $\mdm=0.25~...~2$~GeV from $u,d,s,c$ quarks
  with SM-like and ($B-3L$)-like couplings. We use our modified version
  of \textsc{Herwig}7 for all curves below 2~GeV.}
\label{fig:herwig}
\end{figure}

Most photons and positrons in hadronic processes come from neutral and
charged pion decays, respectively. These pions are either directly
produced or are the end of a decay chain of all forms of hadronic
states listed in Tab.~\ref{tab:channels} in the Appendix. In a few
cases, photons can also be directly produced in DM annihilation, for
instance
\begin{align}
\chi\chi \to \eta\gamma, \pi\gamma \; .
\end{align}
In the left panel of Fig.~\ref{fig:herwig} we see how photon
production channels drop out when we reduce the DM mass or
center-of-mass energy of the non-relativistic scattering
process. Whereas for $\mdm > 1$~GeV all possible hadronic final
states contribute to the round shape of the spectrum, for lower
energies only photons and positrons from very specific processes give
a characteristic energy spectrum. 

For example for $\mdm = 500$~MeV or equivalently a center-of-mass
energy of 1~GeV we expect two kaons from the $\phi$ resonance to
provide most photons through consecutive decays of kaons to pions to
photons. This leads to a triangular shape of the photon spectrum. If
we go down to 250~MeV, the only remaining annihilation channels are
\begin{align}
\chi \chi \to \pi^0\gamma, \pi\pi, 3\pi \; .
\end{align}
Of those, the photons mainly come from the $\pi^0\gamma$ final state,
so one photon is produced directly with an energy around the DM
mass. It leads to the sharp peak around $x \approx 1$. The additional
photons come from the $\pi^0$-decay and are responsible for the
distribution to roughly $x \approx 10^{-1}$. The same applies for
$\mdm = 375$~MeV with an additional bump-like contribution from
neutral pions in the $3\pi$ and $4\pi$ channels as well as additional
photons from the dominantly neutral $\eta\gamma\to
(2\gamma)\gamma,(3\pi^0)\gamma$ decay including a direct
photon.\bigskip

\begin{figure}[t]
\centering
\includegraphics[width=0.49\textwidth]{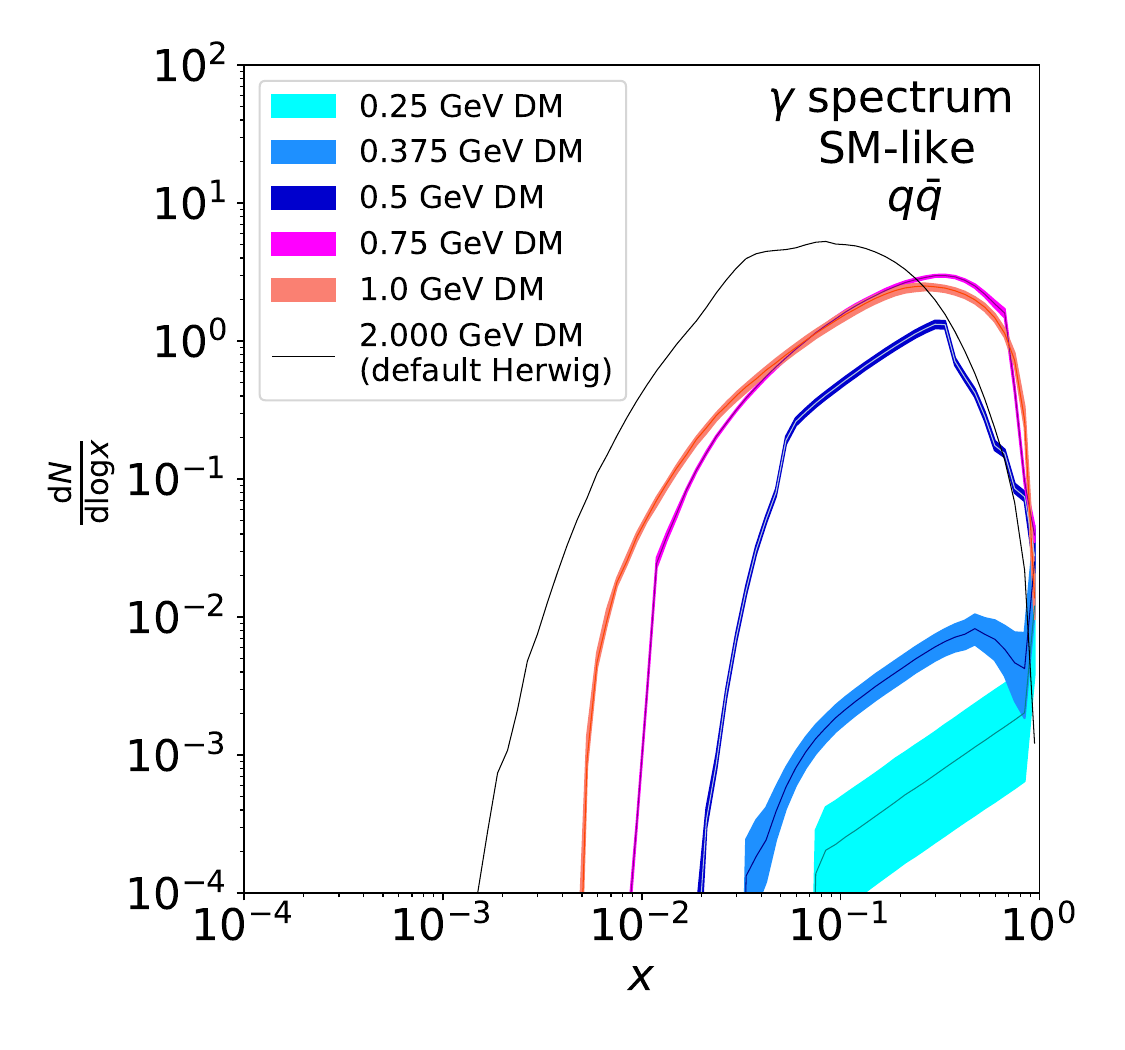}
\includegraphics[width=0.49\textwidth]{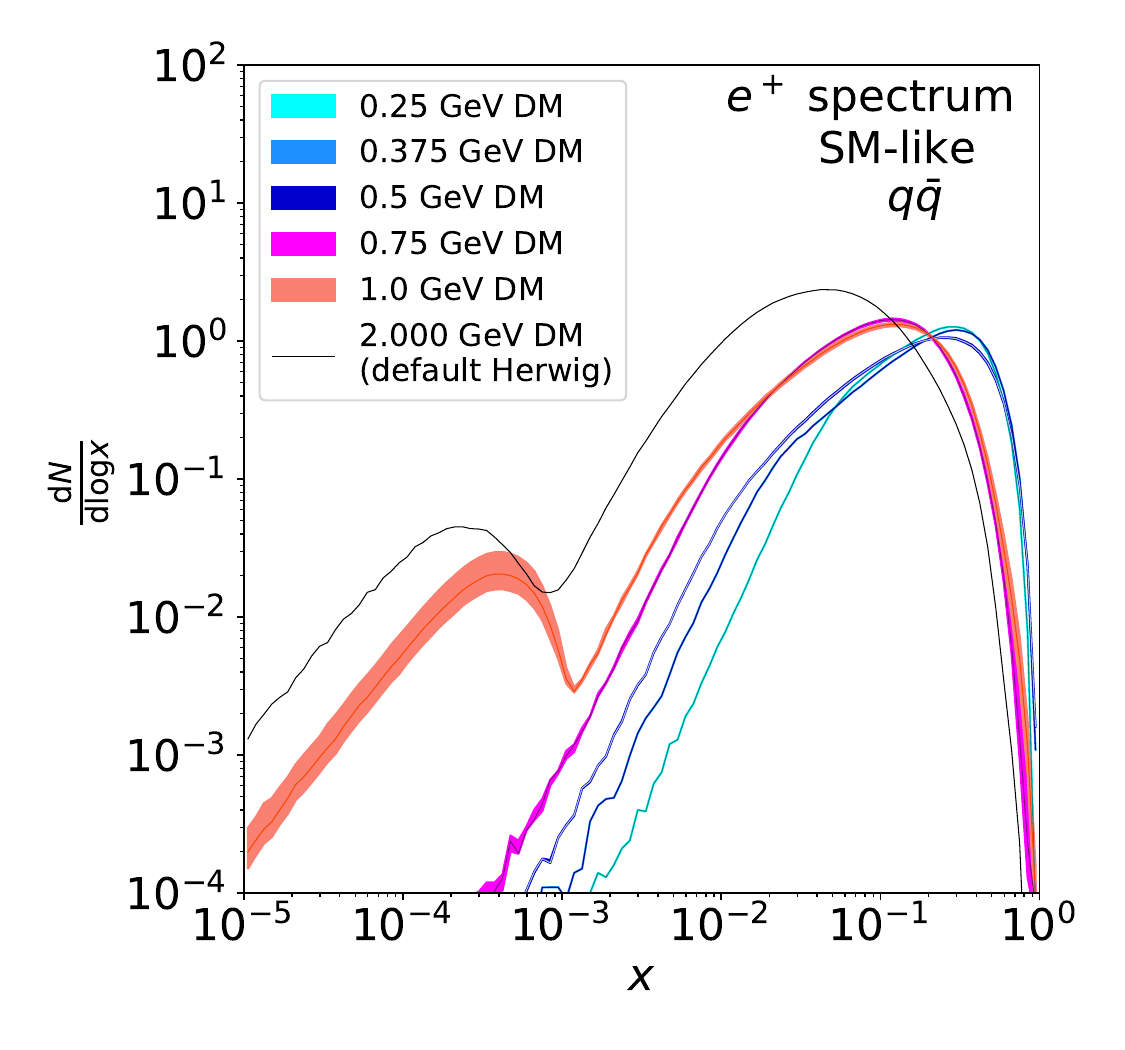}\\
\includegraphics[width=0.49\textwidth]{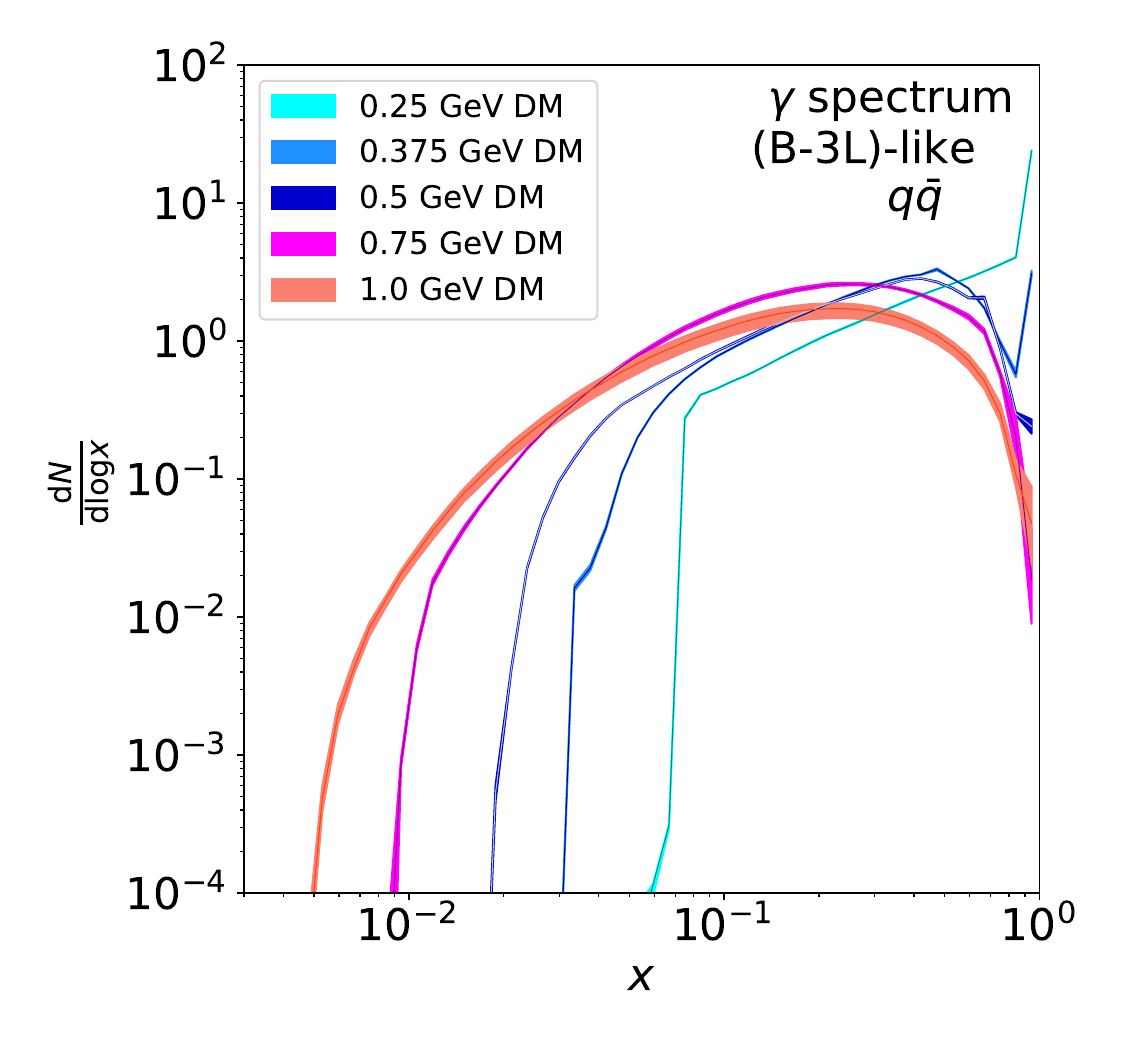} 
\includegraphics[width=0.49\textwidth]{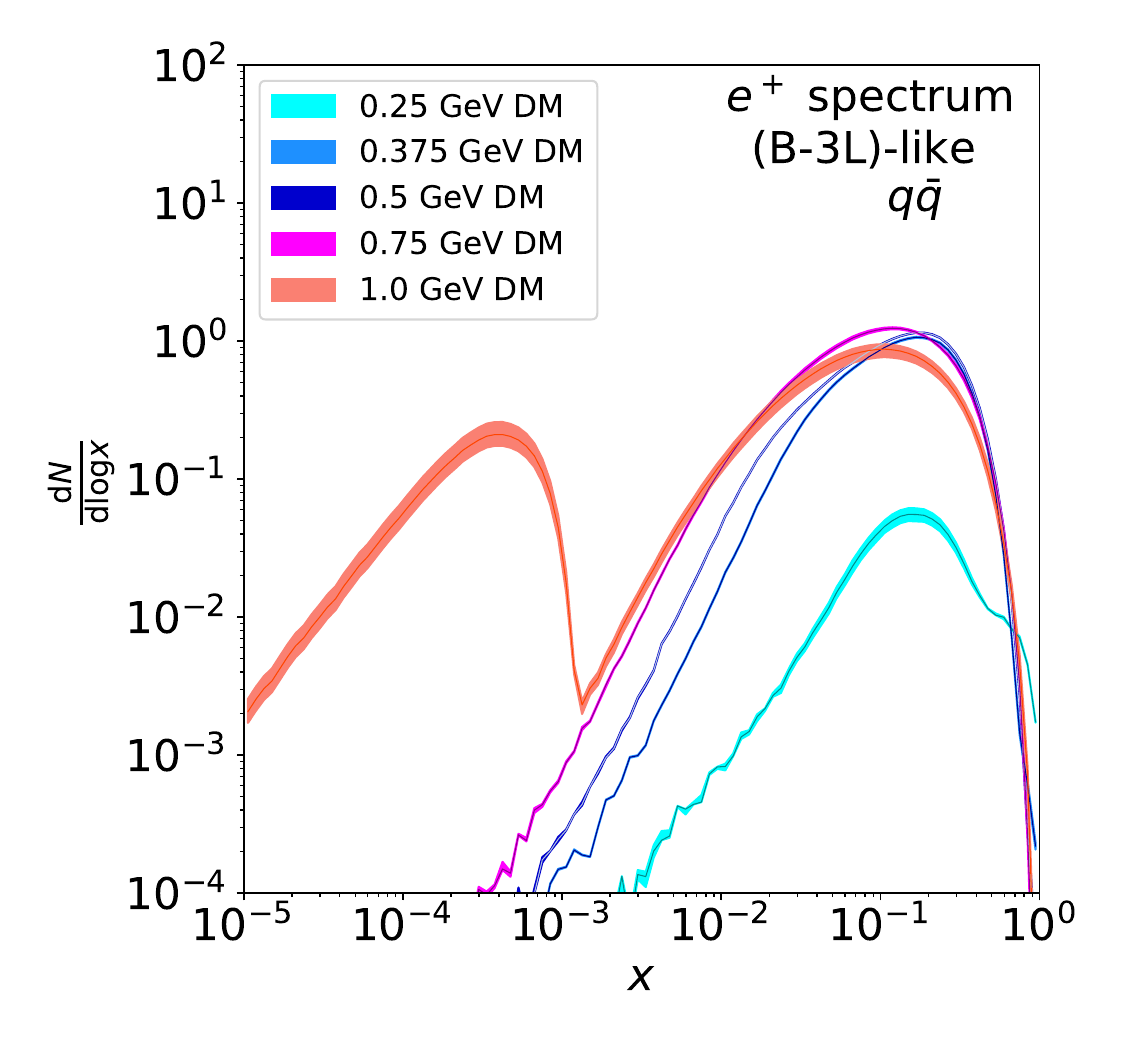}
\caption{Photon and positron spectra $dN /d \log x$ with $x =
  E_\text{kin}/\mdm$ for $\mdm=0.25~...~2$~GeV from $u,d,s,c$ quarks
  with SM-like and ($B-3L$)-like couplings with uncertainty bands
  allowing for perfect cancellations. The 2~GeV curve and the central
  values correspond to Fig.~\ref{fig:herwig}.}
\label{fig:uncertainties}
\end{figure}

The basic shape of the positron spectrum is given by the neutron pair
production threshold. Above threshold, we observe an additional peak
slightly above $x \sim 10^{-4}$ resulting from positron production in
the neutron $\beta$-decay. For $\mdm < 1$~GeV, all positrons come from
charged pion decays. The peak position depends on how early that
charged pion decay occurs for the dominant processes at the respective
center-of-mass energy. For example, for $\mdm < 500$~MeV, charged
pions are mainly produced directly in $\pi\pi, 3\pi, 4\pi$ production
and hence the peak of the spectrum is shifted towards $x=1$.

As mentioned in Sec.~\ref{sec:model}, the composition of the DM
current changes with the way the mediator couples to quarks. In any
($B-3L$)-like model with equal couplings to quarks, the isospin $I=1$
contribution vanishes and consequently some resonance contributions as
well as all channels listed in Eq.\eqref{eq:I1} vanish. For $\mdm =
250$~MeV this implies that without the $\pi\pi$ channel, $\pi^0\gamma$
becomes the dominant annihilation mode. The direct photon production
lifts the photon spectrum, as seen in the upper panels of
Fig.~\ref{fig:herwig}. This is accompanied by a drop in the positron
spectrum that only receives contributions from the subdominant $3\pi$
final state. If we choose a center-of-mass energy below the $3\pi$
threshold, positron spectra from quarks would be completely
absent. For $\mdm = 375$~MeV with an increasing $3\pi$ contribution
towards the $\omega(782)$ resonance, the position spectra are
lifted. For higher energies and the contribution from several
channels, the ($B-3L$)-like spectra resemble the SM-like case. Although
their sources are not identical channel by channel, the way the
photons and positions are produced is similar.

\subsubsection*{Error bands}

\begin{figure}[t]
\centering
\includegraphics[width=0.49\textwidth]{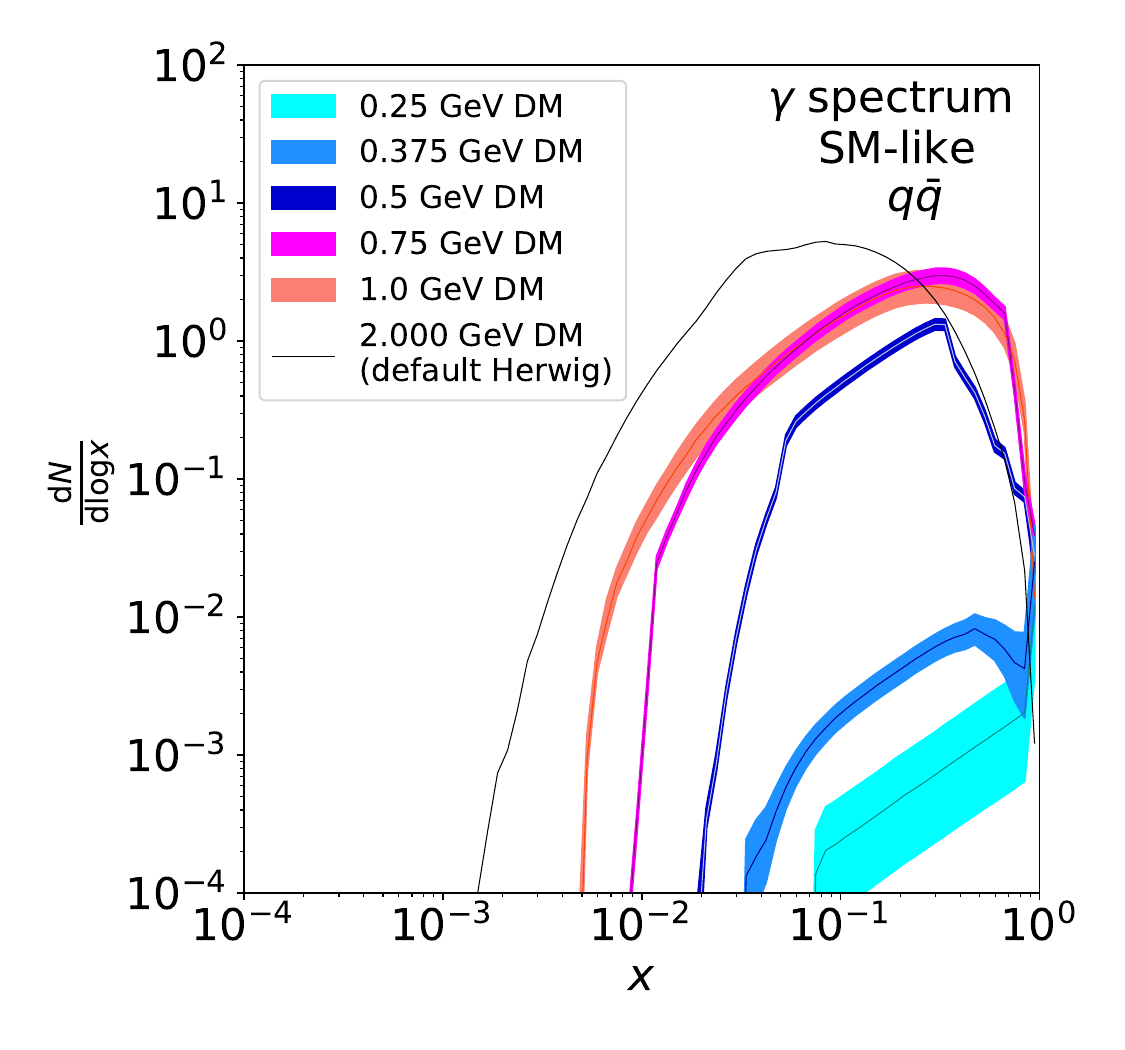}
\includegraphics[width=0.49\textwidth]{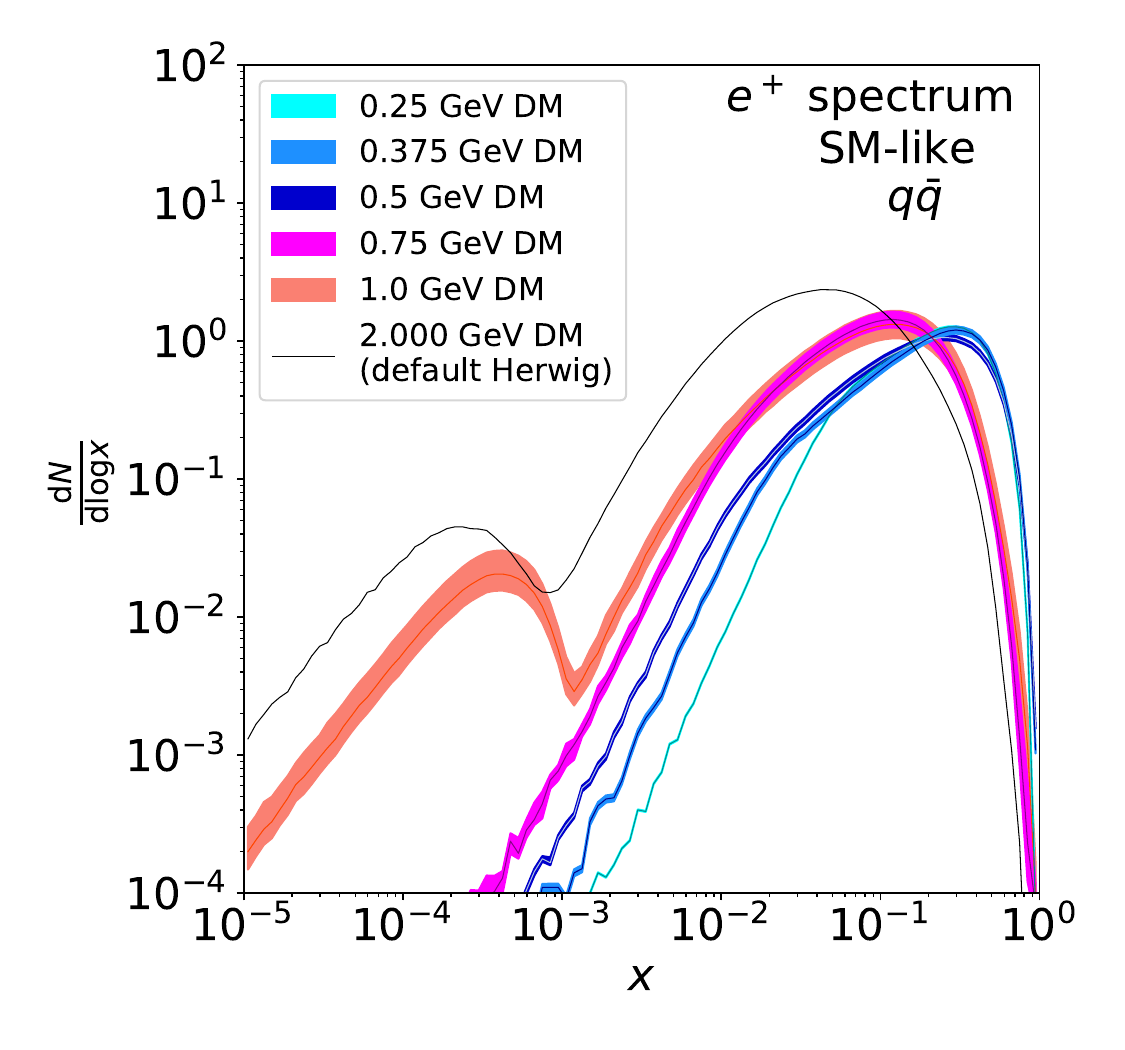}\\
\includegraphics[width=0.49\textwidth]{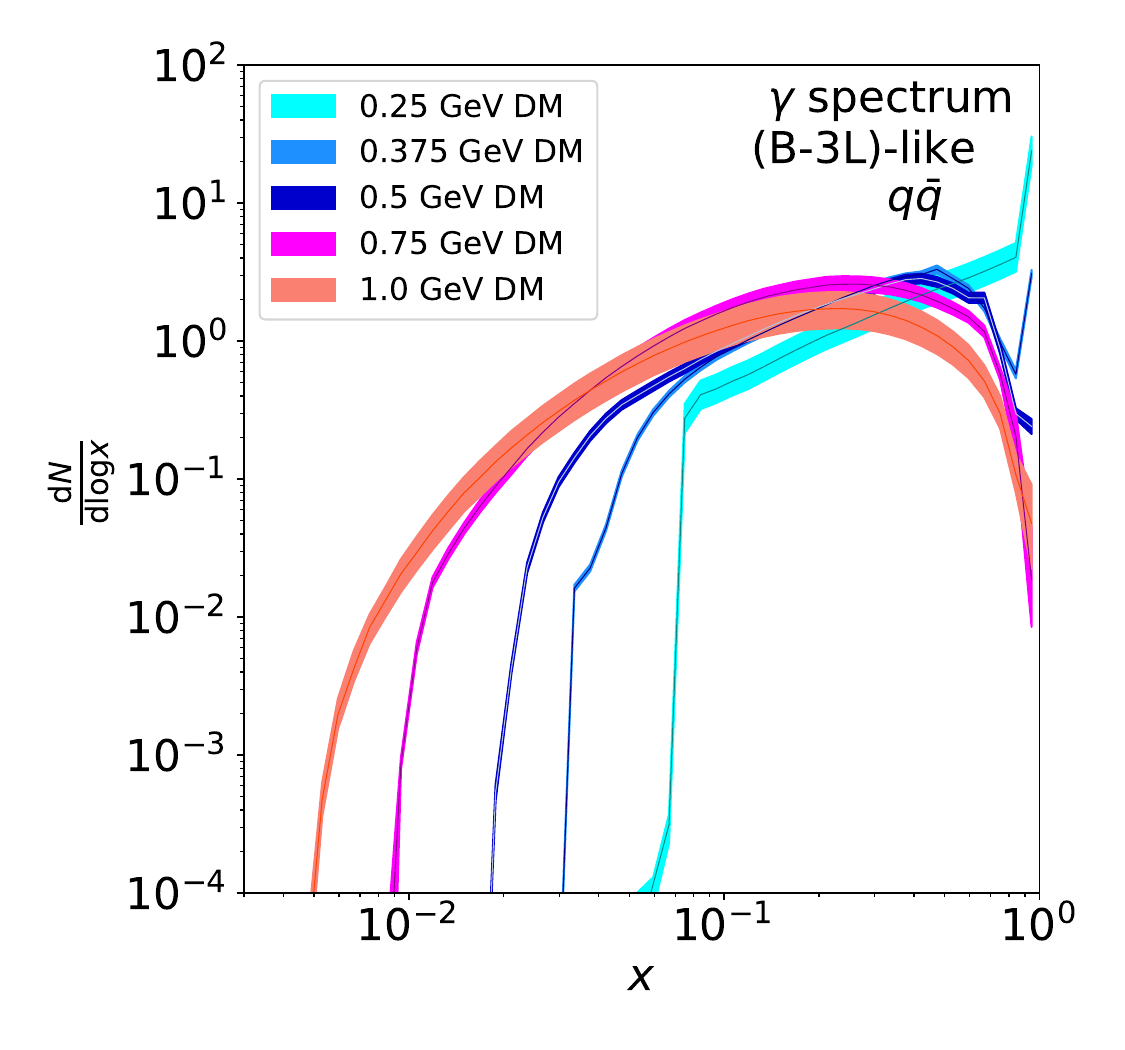} 
\includegraphics[width=0.49\textwidth]{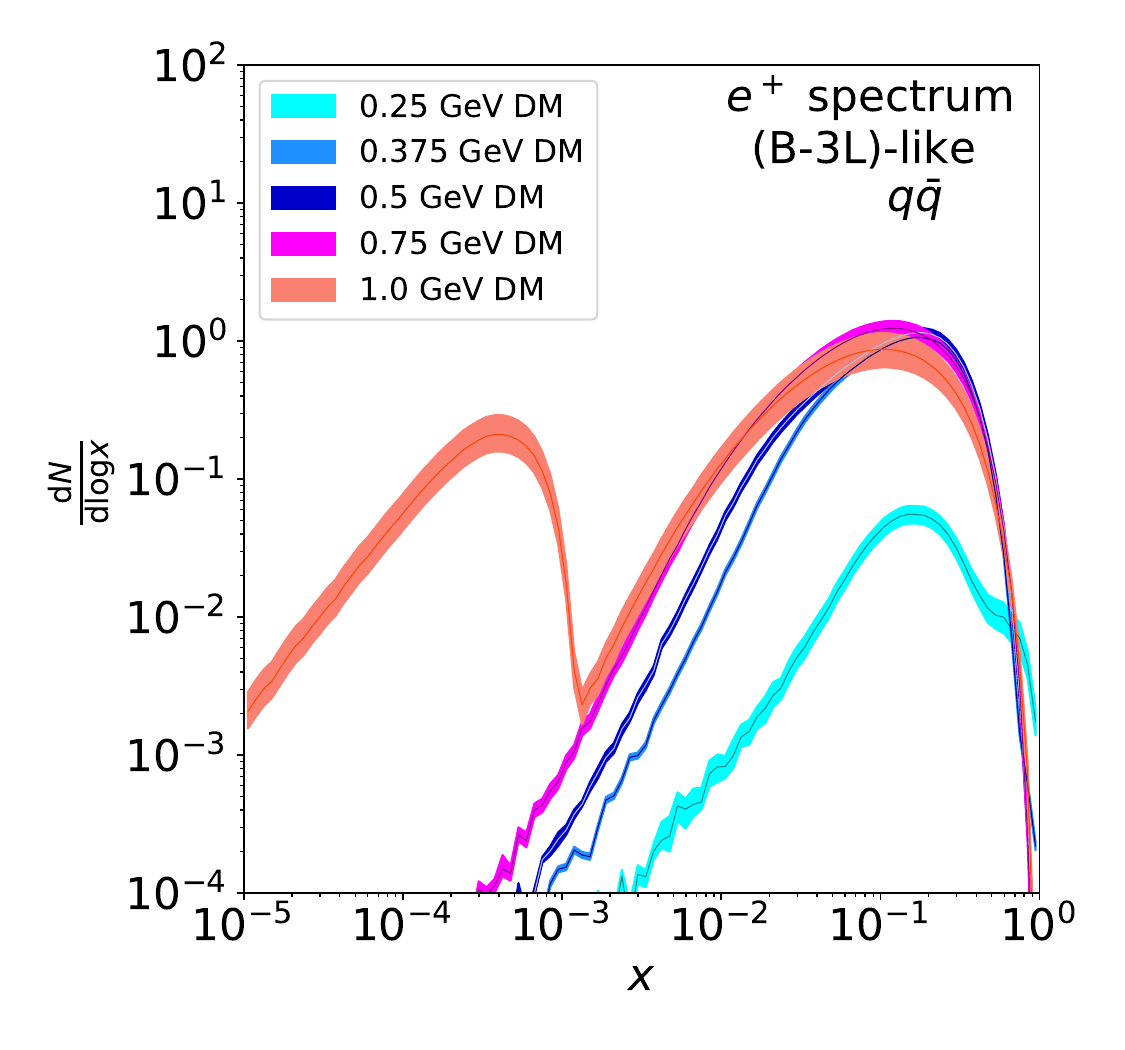}
\caption{Photon and positron spectra $dN /d \log x$ with $x =
  E_\text{kin}/\mdm$ for $\mdm=0.25~...~2$~GeV from $u,d,s,c$ quarks
  with SM-like and ($B-3L$)-like couplings with very conservative
  uncertainty bands. The 2~GeV curve and the central values correspond
  to Fig.~\ref{fig:herwig}.}
\label{fig:uncertainties_cons}
\end{figure}

Uncertainties on the energy spectra are dominated by the uncertainties
from the fits to electron-positron data discussed in the Appendix. We
define ranges of model parameters to cover bands in the
$e^+e^-$-annihilation cross sections as a function of the energy and
propagate those parameter ranges through the hadronic currents into
the energy spectra. This means that the error on a given spectrum
corresponds to the uncertainty of the dominant channel at the
corresponding energy.

In the upper panels of Fig.~\ref{fig:uncertainties} we see that the
photon spectrum at $\mdm =250$~MeV inherits large uncertainties from
the poorly measured dominant $\pi^0\gamma$ channel in that energy
range. For $\mdm = 375$~MeV the more precisely measured $3\pi$ channel
suppresses the $\pi^0\gamma$ channel, but still leaves us with visible
error bands. For even higher energies several channels contribute to
the uncertainty of the photon spectrum. We observe the smallest error
bands for spectra that benefit from precisely measured dominant
processes, for instance peak regions such as the $\phi$ resonance at
1~GeV in the $KK$ channel, the $\rho$ resonance in the $2\pi$ decay,
or generally well-measured channels such as $4\pi$. Positron spectra with
their dominant $2\pi,3\pi,4\pi$ channels are always well measured.
The only exception is $\mdm = 1$~GeV spectrum, especially the lower
peak around $\sim 10^{-4}$ , which comes from the neutron
$\beta$-decay. As discussed in the Appendix, the $n\bar{n}$ channel is
poorly measured and leaves us with larger uncertainties in that
regime.

In ($B-3L$)-like models, we will not get any contributions from
well-measured $2\pi$ and $4\pi$ final states. This means the
uncertainties on the position spectrum for $\mdm = 250$~MeV are
slightly larger than in the SM-like case, see the lower panel of
Fig~\ref{fig:uncertainties}. Nevertheless, as long as no channel drops
out and another channel with larger uncertainties starts to dominate,
the uncertainties in the ($B-3L$)-like case tend to be smaller. The reason is
the absence of the $I=1$ contributions and their sizeable
uncertainties.\bigskip

Finally, we want to ensure that our error estimates are
conservative. In Fig.~\ref{fig:uncertainties} we use the uncertainties
on the individual channels bin-wise, add all contributions up and
normalize by the sum of their corresponding cross-sections. For
channels with large cross-sections that are also giving the main
contribution to the total amount of photons/positrons in the spectrum,
the error bars can completely cancel for the normalized spectra. This
way, we only get sizable uncertainty bands for spectra where one
channel is dominating the shape of the spectrum, but is playing a
sub-dominant role in the total cross-section. An example is the
$\pi^0\gamma$ final state for the SM-like photon spectrum at $\mdm =
250$~MeV or the lower bump in the 1~GeV positron spectrum caused by
$n\bar{n}$. This assumption can be considered somewhat aggressive in a
situation where we do not have full control of the full error
budget. Instead, we can maximize and minimize all spectra channel by
channel and separately normalize them by the smallest and largest
total cross-section possible. This way there will be no cancellation
for single-channel spectra, and in Fig.~\ref{fig:uncertainties_cons}
we indeed see much increased uncertainties. Obviously, the real error
bands are going to be somewhere between the results shown in
Fig.~\ref{fig:uncertainties} and Fig.~\ref{fig:uncertainties_cons}
determined by analysis details beyond the scope of this first
analysis.

\section{Outlook}

We have studied the positron and photon spectra from non-relativistic
dark matter annihilation in a dark matter mass range from 250~MeV to
5~GeV (with the exception of the poorly understood region near the charm threshold).
We consider a light vector mediator with general
couplings to SM fermions. For the photon spectra we see a smooth
interpolation from typical hadron decay chains with their round
spectra down to the pion continuum with a triangular shape. For
positrons the main feature is the secondary neutron decay above
threshold.

Because we are relying on an updated fit to electron-position input
data to \textsc{Herwig} we can also propagate the uncertainties from
poorly measured channels into the photon and positron spectra. Already
for relatively heavy dark matter the positron spectrum shows sizeable
error bars. In the case of photons, smaller dark matter masses with
fewer and less well measured annihilation channels are also plagued by
significant error bars, eventually covering an order of magnitude for
$\mdm = 250$~MeV.

Our new implementation closes the gap between standard
\textsc{Pythia}-based tools such as \textsc{Pppc4dmid},
\textsc{MicrOMEGAS}, \textsc{MadMD}, or \textsc{DarkSUSY} and the
comparably simple small-mass continuum regime and should allow for a
reliable study of GeV-scale dark matter even if it dominantly
interacts with SM quarks.\footnote{The code we have used to produce
  these results will be available in a future version of
  \textsc{Herwig}7.  If there is sufficient interest we will also
  think about providing the output as a cool and fast neural network.}
\newline
\newline

\begin{center} \textbf{Acknowledgments} \end{center}

First, we want to thank Patrick Foldenauer for his early contributions
to Figs.~\ref{fig:5gev} and~\ref{fig:2gev}. TP is supported by the
German Research Foundation DFG under grant no.~396021762-TRR~257 and
would like to thank Stefan Gieseke for help starting this project. Peter Reimitz
is funded by the Graduiertenkolleg \textit{Particle physics beyond the
  Standard Model} (GRK 1940). Peter Richardson
is supported by funding from the UK Science and Technology Facilities Council
(grant numbers ST/P000800/1, ST/P001246/1), and benefited from
the European Union’s Horizon 2020 research and innovation 
programme as part of the Marie Sk\l{}odowska-Curie Innovative Training Network MCnetITN3
(grant agreement no. 722104).

\clearpage
\appendix
\section{Updated fits with error envelopes}

\begin{table}[t!]
\centering
\begin{tabular}{l|lccc}
\hline
Channel & Data & Parametrization & fit & threshold [GeV]\\ \toprule
$\pi\gamma$ & \cite{Achasov:2016bfr} &\cite{Achasov:2016bfr} &
                                                               \cite{Achasov:2016bfr}
                                       &  \\ \midrule
$\pi\pi$ & \cite{Ambrosino:2008aa,Aubert:2009ad,Lees:2012cj} &
                                                               \cite{Czyz:2010hj}
                                 &\cite{Czyz:2010hj}  & 0.280\\
$\pi\pi\pi$ & \cite{Aubert:2004kj}& \cite{Czyz:2005as} &
                                                         \cite{Czyz:2005as} & 0.420\\
$4\pi$ & \cite{Lees:2012cr,TheBaBar:2017vzo} & \cite{Czyz:2008kw} &
                                                                      own
                                       & 0.560\\
$\omega\pi$ & \cite{Achasov:2016zvn} & \cite{Achasov:2016zvn} &
                                                                \cite{Achasov:2016zvn} & 0.918\\
$p\bar{p}/n\bar{n}$ &
                      \cite{Lees:2013ebn,Ablikim:2015vga,Pedlar:2005sj,Delcourt:1979ed,Castellano:1973wh,Antonelli:1998fv,Armstrong:1992wq,Ambrogiani:1999bh,Andreotti:2003bt,Punjabi:2005wq,Puckett:2011xg,Gayou:2001qt,Puckett:2010ac,Puckett:2017flj,Pospischil:2001pp,Plaster:2005cx,Geis:2008aa,Andivahis:1994rq}
               & \cite{Czyz:2014sha} & own & 1.877\\
$\eta\gamma$ & \cite{Achasov:2006dv} & \cite{Achasov:2006dv} &
                                                               \cite{Achasov:2006dv}
                                       & 0.548\\
$\eta\pi\pi$ & \cite{Achasov:2017kqm,TheBABAR:2018vvb}&
                                                        \cite{Czyz:2013xga}
            & own & 0.827\\
$\eta' \pi\pi$ & \cite{Aubert:2007ef} & \cite{Czyz:2013xga} & own & 1.237\\
$\omega\pi\pi$ & \cite{Akhmetshin:2000wv,Aubert:2007ef,Lees:2018dnv}&
                                                                      own
            & own & 1.062\\
$\eta\phi$  & \cite{Aubert:2007ym,Achasov:2018ygm}& own & own & 1.568\\
$\eta\omega$ &\cite{Achasov:2016qvd} & own & own & 1.331\\
$\phi\pi$ & \cite{Aubert:2007ym,TheBABAR:2017vgl}& own & own & 1.160\\
$KK$ &
       \cite{Achasov:2000am,Achasov:2006bv,Mane:1980ep,Kozyrev:2016raz,Lees:2014xsh,Pedlar:2005sj,Akhmetshin:2008gz,Lees:2013gzt,Lees:2015iba,Achasov:2016lbc}
               & \cite{Czyz:2010hj} & own & 0.996\\
$KK\pi$ &
          \cite{TheBABAR:2017vgl,Aubert:2007ym,Achasov:2017vaq,Bisello:1991kd,Mane:1982si}&
                                                                                            own
                                 &own & 1.135\\
\bottomrule
\end{tabular}
\caption{Dominant processes contributing to $e^+e^- \to$~hadrons
  in the relevant energy range.}
\label{tab:channels}
\end{table}

If we limit ourselves to dark matter annihilation through a vector
mediator we can relate the dark matter annihilation process to the
corresponding and measurable process
\begin{align}
e^+e^- \to \text{hadrons.}
\end{align}
Its matrix element has the form
\begin{align}
\mathcal{M} = \frac{e}{\hat{s}} \; 
\bar{v}_{e^+} \gamma_\mu u_{e^-} \; 
\langle \text{had}|J_\text{em}^\mu|0\rangle \; .
\end{align}
The electromagnetic quark current $J_\text{em}^\mu=\sum_{q=u,d,s}
e_q \bar{q}\gamma^\mu q$ can be decomposed into its isospin 
components $I=0,1$ and its strange-quark content,
\begin{align}
J_\text{em}^\mu 
=\frac{1}{\sqrt{2}}J_{I=1,3}^\mu
+\frac{1}{3\sqrt{2}}J_{I=0}^\mu
-\frac{1}{3}J_s^\mu \; ,
\end{align}
with
\begin{align}
J_{I=1,3}^\mu &= \frac{\bar{u}\gamma_\mu u-\bar{d}\gamma_\mu d}{\sqrt{2}}, \notag \\
J_{I=0}^\mu &= \frac{\bar{u}\gamma_\mu u+\bar{d}\gamma_\mu d}{\sqrt{2}}, \notag \\
J_s^\mu &= \bar{s}\gamma_\mu s \; .
\end{align}
We study all hadronic states which appear in the total cross section
$\sigma(e^+e^-\to \text{hadrons})$ in the MeV to GeV range. A list of
all channels, their parametrizations, their data fits, and their
threshold values is given in Tab.~\ref{tab:channels}. Our modelling of
the $e^+e^-$ scattering relies on vector meson
dominance~\cite{OConnell:1995nse}. In that case the hadronic current
$\langle \text{had}|J_\text{em}^\mu|0\rangle$ can be described by a
momentum-dependence and a form-factor that includes all resonances
allowed under certain isospin symmetry assumptions. The
parametrization and fit values for the form-factors for the
$\pi\gamma$, $\pi\pi$, $3\pi$, $\omega\pi$, and $\eta\gamma$ final
states are taken from
Refs.~\cite{Czyz:2010hj,Czyz:2005as,Czyz:2008kw}, as implemented in
the event generator
\textsc{Phokhara}~\cite{Rodrigo:2001kf,Czyz:2017veo}, and the Born cross section
formulae from the SND
measurements~\cite{Achasov:2016bfr,Achasov:2016zvn,Achasov:2006dv}. For
all other channels, we provide new fits. Our modelling does not take
into account possible final state interactions such as rescattering~\cite{Cohen:1980cq}
and Sommerfeld-effects of non-relativistic final states~\cite{Cassel:2009wt}. For example,
the K-matrix approach~\cite{Aitchison:1972ay} includes such interactions, \eg the $\pi\pi
\leftrightarrow KK$ rescattering above the $KK$ threshold, with an infinite series of
rescattering loops. It is used to describe, for example, three-body $B$-decays~\cite{PhysRevD.92.054010}. The
only exception of using rescattering effects is the Flatt\'{e}
parametrization in the $\omega\pi\pi$ channel that takes into account
$KK$ threshold effects as seen below.

\subsubsection*{\boldmath$p\bar{p}$ (update)}

The data and the fit function for this channel are given in
Tab.~\ref{tab:channels}. We updated the data set used for our fit
since from the input to the previous fit~\cite{Czyz:2014sha}
Ref.~\cite{Gayou:2001qd} is superseded by Ref.~\cite{Puckett:2011xg},
Ref.~\cite{Madey:2003av} by Ref.~\cite{Plaster:2005cx}, and
Ref.~\cite{Ablikim:2005nn} by Ref.~\cite{Ablikim:2015vga}. For
asymmetric data uncertainties we symmetrize statistical and systematic
uncertainties separately and then add both in quadrature. We refrain
from a more sophisticated error analysis for instance including
correlations between systematic uncertainties, since in most cases
detailed information about the systematic uncertainties is either
missing or the statistical uncertainty dominates. For the fit, we get
$\chi^2/\text{n.d.f} = 1.069$, and the best-fit values are shown in
Tab.~\ref{tab:ppbar}.

\begin{table}[t]
\centering
\begin{tabular}{cc|cc|cc|cc}
\toprule
$c_1^{1R}$   & -0.467(12)  & $c_1^{1I}$   & -0.385(15) &
$c_2^{1R}$   &-0.177(11) & $c_2^{1I}$   & 0.149(12) \\
$c_3^{1R}$   &  0.301(18)& $c_3^{1I}$   & 0.264(16)  &
$c_1^{2R}$   &  0.052(13) & $c_1^{2I}$   &  -3.040(21)\\
$c_2^{2R}$   &  -0.003(11) & $c_2^{2I}$  &  2.380(15)&
$c_3^{2R}$   &  -0.348(11) & $c_3^{2I}$   &  -0.104(12)\\
$c_1^{3R}$   &  -7.88(47) & $c_1^{3I}$  &  5.67(29)&
$c_2^{3R}$   &  10.20(10) & $c_2^{3I}$   &  -1.94(31)\\
$c_1^{4R}$   & -0.8320(11)  & $c_1^{4I}$  &  0.3080(12)&
$c_2^{4R}$   &  0.4050(11) & $c_2^{4I}$   &  -0.2500(12)\\
\bottomrule
\end{tabular}
\caption{Parameters of the nucleon form factor from our fit using the model
  describing $pp$ production from Ref.~\cite{Czyz:2014sha}.}
\label{tab:ppbar}
\end{table}

\subsubsection*{\boldmath$\eta\pi\pi, \eta'\pi\pi$ (update)}

The fit function for the $\eta\pi\pi$ and $\eta'\pi\pi$ hadronic
currents are based on \cite{Czyz:2013xga}. We re-fit the fit function
to more recent data sets \cite{Achasov:2017kqm,TheBABAR:2018vvb}
compared to those used in \cite{Czyz:2013xga}. The fit values can be
found in Tab.~\ref{tab:etapipi}. 
%

\begin{table}[b!]
\centering
\begin{tabular}{lrr|lrr}
\toprule
Parameter & $\eta\pi\pi$ & $\eta'\pi\pi$ &
Parameter & $\eta\pi\pi$ & $\eta'\pi\pi$\\ \midrule
$m_{\rho_1}$ [GeV] & 1.5400(39) & - &
$a_{1}$ & 0.326(10)& 0 (fixed)\\ 
$m_{\rho_2}$ [GeV]& 1.7600(58) & - &
$a_2$ & 0.0115(31)& 0 (fixed)\\ 
$m_{\rho_3}$ [GeV] & 2.15 (fixed)& 2.110(36) &
$a_3$ & 0 (fixed)& 0.0200(81) \\
$\Gamma_{\rho_1}$ [GeV] & 0.356(17)&  - &
$\varphi_1$ & $\pi$ (fixed)& -\\
$\Gamma_{\rho_2}$ [GeV] & 0.113(22)& -  &
$\varphi_2$ & $\pi$ (fixed)& - \\
$\Gamma_{\rho_3}$ [GeV]  & 0.32 (fixed)& 0.18(11) &
$\varphi_3$ & 0 (fixed) & $\pi$ (fixed) \\
&&& $\chi^2/\text{n.d.f}$ & 0.8732 & 0.9265\\
\bottomrule
\end{tabular}
\caption{Fit values for the $\eta\pi\pi$ and $\eta'\pi\pi$ channels.}
\label{tab:etapipi}
\end{table}

\subsubsection*{\boldmath$KK$ (update)}

We parametrize the hadronic current for the $K^0\bar{K}^0$ and
$K^+K^-$ channels in the same way as done in
Ref.~\cite{Czyz:2010hj}. Unlike Ref.~\cite{Czyz:2010hj}, we do not fix
all masses and widths of the $\rho, \omega$ and $\phi$ states to their
PDG values but let them float in the fit. Furthermore, we use an
updated data set for the fit, as mentioned in
Tab.~\ref{tab:channels} and included the $\tau^-\to K^0_S\pi^-\nu_\tau$
data from Ref. \cite{Epifanov:2007rf} to better constrain the $I=1$ component of the current.
The fit values are listed in
Tab.~\ref{tab:KK}. The last coupling of each resonance is calculated
via Eq.(16) in Ref.~\cite{Czyz:2010hj}, and we keep
$\eta_{\phi}=1.055$, $\gamma_{\omega}=0.5$ and $\gamma_{\phi}=0.2$
fixed such as in Ref.~\cite{Czyz:2010hj}. For the simultaneous fit to
$K^0K^0$ and $K^+K^-$ data we obtain $\chi^2/\text{n.d.f}=1.621$.

\begin{table}[t]
\centering
\begin{tabular}{cc|cc|cc|cc}
\toprule
$m_{\rho_0}$   & 0.77549 (PDG)  & $\Gamma_{\rho_0}$   & 0.1494 (PDG)&
$c_{\rho_0}$   & 1.1149(24) & $c_{\rho_4}$   & -0.0383(66)\\
$m_{\rho_1}$   & 1.5207(53) & $\Gamma_{\rho_1}$   & 0.213(14)&
$c_{\rho_1}$   & -0.0504(44) & $c_{\rho_5}$   & 0.0775 (calc.) \\
$m_{\rho_2}$   & 1.7410(38) & $\Gamma_{\rho_2}$   & 0.084(12)&
$c_{\rho_2}$   & -0.0149(32) & $\beta_{\rho}$   & 2.1968 \\
$m_{\rho_3}$   & 1.992(15) & $\Gamma_{\rho_3}$   & 0.290(41)&
$c_{\rho_3}$   & -0.0390(45) &- & -\\
$m_{\omega_0}$   & 0.78265 (PDG) & $\Gamma_{\omega_0}$   & 0.00849 (PDG) &
$c_{\omega_0}$   & 1.365(44) & $c_{\omega_3}$   & 1.40(27) \\
$m_{\omega_1}$   & 1.4144(71)  & $\Gamma_{\omega_1}$   & 0.0854(71)&
$c_{\omega_1}$   & -0.0278(83)& $c_{\omega_4}$   & 2.8046 (calc.) \\
$m_{\omega_2}$   & 1.6553(26)  & $\Gamma_{\omega_2}$   & 0.1603(26)&
$c_{\omega_2}$   & -0.325(30) & $\beta_{\omega}$   & 2.6936\\
$m_{\phi_0}$   & 1.0194209(94)  & $\Gamma_{\phi_0}$   & 0.004253(21)&
$c_{\phi_0}$   & 0.9658(27) & $c_{\phi_3}$   & 0.1653(50) \\
$m_{\phi_1}$   & 1.5948(51) & $\Gamma_{\phi_1}$   & 0.029(18)&
$c_{\phi_1}$   & -0.0024(20) & $c_{\phi_4}$   & 0.1195 (calc.) \\
$m_{\phi_2}$   & 2.157(57)  & $\Gamma_{\phi_2}$   & 0.67(16) &
$c_{\phi_2}$   & -0.1956(19) & $\beta_{\phi}$  & 1.9452\\
\bottomrule
\end{tabular}
\caption{Parameters for the description of $KK$ production from our fit using the model of
  Ref.~\cite{Czyz:2010hj}. All masses and widths are given in GeV, all
other parameters are dimensionless}
\label{tab:KK}
\end{table}

\subsubsection*{\boldmath$4 \pi$ (update)}

For the $4\pi$ channel, we use the parametrization of
Ref.~\cite{Czyz:2008kw} and fit it to more recent rate measurements
for $e^+e^- \to 2\pi^0\pi^+\pi^-$ and $e^+e^- \to 2\pi^+2\pi^-$ from
BaBar~\cite{Lees:2012cr,TheBaBar:2017vzo}. We obtain a
$\chi^2/\text{n.d.f} = 1.28$ and the fit values are listed in
Tab.~\ref{tab:4pi}.

\begin{table}[b!]
\centering
\begin{tabular}{l|cc|l|cc}
\toprule
$\bar{m}_{\rho_1}$   & 1.44 (fixed) & $\bar{m}_{\rho_2}$   & 1.74 (fixed) &
$\bar{m}_{\rho_3}$   & 2.12 (fixed) \\
\hline
$\bar{\Gamma}_{\rho_1}$   &  0.678(18)&$\bar{\Gamma}_{\rho_2}$  & 0.805(29)  &
$\bar{\Gamma}_{\rho_3}$   &  0.209(29) \\
\hline
$\beta_1^{a_1}$   &  -0.0519(56) & $\beta_2^{a_1}$  &  -0.0416(20) &
$\beta_3^{a_1}$   & -0.00189(47) \\
\hline
$\beta_1^{f_0}$   &  $7.39(0.29)\cdot 10^4$ & $\beta_2^{f_0}$  &
                                                                 $-2.62(0.19)
                                                                          \cdot
                                                                 10^3$&
$\beta_3^{f_0}$   &  334(87) \\
\hline
$\beta_1^\omega$   & -0.367(27)  & $\beta_2^\omega$  &  0.036(11)&
$\beta_3^\omega$   &  -0.00472(77) \\
\hline
$c_{a_1}$   &-202.0(24) & $c_{f_0}$  &  124.0(52)&
$c_\omega$   &  -1.580(73) \\
\hline
$c_\rho$ & -2.31(24) & $\chi^2$ & 291 & n.d.f & 228\\
\bottomrule
\end{tabular}
\caption{Parameters for the $4\pi$ channel for our fit using the model
  from \cite{Czyz:2008kw}. All masses and widths
 are in GeV; couplings $\beta_i^j$, ($j=a_1,f_0,\omega$ and $i=1,2,3$)
as well as $c_\rho$ are dimensionless; $c_{a_1}$ and $c_{f_0}$ in
$\text{GeV}^{-2}$ and $c_\omega$ in $\text{GeV}^{-1}$.}
\label{tab:4pi}
\end{table}

\subsubsection*{\boldmath$\eta\phi,\eta\omega,\phi\pi$ (new)}

Our first new fit is to the processes $e^+
e^- \to \eta\phi,\eta\omega,\phi\pi$, where the momentum-dependent
Born cross sections are
\begin{align}
\sigma(s) = \frac{4\pi\alpha_\text{em}(s)^2}{3\hat{s}^{3/2}} \; P_f(s) \; |F|^2,
\end{align}
where $\alpha_\text{em}(s)$ is the fine structure constant,
$P_f(s)=q_{\text{cm},X}^3$ the final-state phase space,
$q_{\text{cm},X}$ the final-state particle momentum and $F$ is the
respective form factor. The resonant contributions are simply
parametrized by
\begin{align}
F_{\eta\omega,\eta\phi} &= \sum_i \frac{a_ie^{i\varphi_i}}{m_i^2-\hat{s}-i m_i\Gamma_i},
\notag \\
F_{\phi\pi} =& \sum_i\frac{a_ie^{i\varphi_i}}{m_i^2-\hat{s}-i\sqrt{\hat{s}}\Gamma(\hat{s})} \; ,
\end{align}
where we take the $s$-dependent width $\Gamma(s)$ from
Ref.~\cite{Aubert:2007ym}. All parameters and fit values for
$\eta\phi$, $\eta\omega$, and $\phi\pi$ production are listed in
Tab.~\ref{tab:fits1}.

\begin{table}[t]
\centering
\begin{small} \begin{tabular}{l|cc|cc|cc}
\toprule
Process & \multicolumn{2}{c!}{$\eta\phi$} & \multicolumn{2}{c|}{$\eta\omega$} & \multicolumn{2}{c}{$\phi\pi$}\\
\hline
$i$ & $\phi'$ & $\phi''$ & $\omega'$ & $\omega''$ & $\rho$ & $\rho'$\\
\hline
$m_i$ [GeV] & $1.67\pm 0.0063$& $2.14\pm 0.012$ & 1.425 \cite{Tanabashi:2018oca}&
                                                                    $1.67\pm
                                                                 0.0087$
              & 0.77526 \cite{Tanabashi:2018oca}& 1.593 \cite{Aubert:2007ym} \\
\hline
$\Gamma_i$ [GeV] & $0.122\pm 0.0075$ & $0.044\pm 0.033$ & 0.215
                                                            \cite{Tanabashi:2018oca} &
                                                                      $0.113\pm
                                                                      0.016$
              & 0.1491 \cite{Tanabashi:2018oca} & 0.203 \cite{Aubert:2007ym} \\
\hline
$a_i$ & $0.175\pm 0.0084$ & $0.0041\pm 0.0019$& $0.0862 \pm 0.011$&
                                                                      $0.0648
                                                                      \pm
              0.0078$ & $0.194\pm 0.073$ & $0.0214\pm 0.0035$\\
\hline
$\varphi_i$ & 0 (fixed) & $2.19\pm 0.046$ & 0 \cite{Achasov:2016qvd} & $\pi$
                                                         \cite{Achasov:2016qvd}&  0
                                                                  (fixed)&
  $121\pm 16.9$ deg.\\
\hline
$\chi^2/\text{n.d.f}$ & \multicolumn{2}{c|}{0.9388} &\multicolumn{2}{c|}{1.3332} & \multicolumn{2}{c}{0.9798}\\
\bottomrule
\end{tabular} \end{small}
\caption{Fit values for the $\eta\phi$, $\eta\omega$, and $\phi\pi$ channels.}
\label{tab:fits1}
\end{table}

\subsubsection*{\boldmath$\omega \pi \pi$ (new)}

Next, for the $\omega\pi\pi$ channel, we use
\begin{align}
\langle \omega\pi\pi | J_\text{em}^\mu|0\rangle= e g^{\mu\nu}
\frac{g_{\omega''}
  m_{\omega''}^2}{\hat{s}-m_{\omega''}^2+im_{\omega''}\Gamma_{\omega''}}g_{\nu\sigma}\varepsilon_\omega^\sigma
\sum_{i=1,2} \text{BW}_{f_i}(q^2)
\end{align}
for the hadronic current. In our energy range we only need to consider
one vector meson mediator $\omega''$, namely the $\omega(1650)$
meson. For the $f_i$ mediator we have
\begin{align}
\text{BW}_{f_1}(m_{\pi\pi})=\frac{g_{\omega''\omega \sigma}m_{\sigma}^2}{m_{\pi\pi}^2-m_{\sigma}^2+im_{\sigma}\Gamma_{\sigma}}
\end{align}
where $m_\sigma$ and $\Gamma_{\sigma}$ are the mass and width of the
$\sigma$ meson and using the Flatt\'{e} parametrization~\cite{Flatte:1976xu}
\begin{align}
\text{BW}_{f_0}(m_{\pi\pi})=\frac{g_{\omega''\omega f_0(980)}m_{f_0(980)}\sqrt{\Gamma_0\Gamma_{\pi\pi}}}{m_{\pi\pi}^2-m_{f_0(980)}^2+im_{f_0(980)}(\Gamma_{\pi\pi}+\Gamma_{\bar{K}K}^*)}
\end{align}
with
\begin{align}
\Gamma_{\pi\pi}&=g_{\pi\pi}q_{\pi}(m_{\pi\pi})\nonumber \\
\Gamma_{\bar{K}K}&=\begin{cases}
      g_{\bar{K}K}\sqrt{(1/4)m_{\pi\pi}^2-m_{K}^2}, & \text{above threshold} \\
      ig_{\bar{K}K}\sqrt{m_K^2-(1/4)m_{\pi\pi}^2}, & \text{below threshold}
    \end{cases}\nonumber \\
\Gamma_{\bar{K}K}^*&=0.5\cdot(\Gamma_{\bar{K}^0K^0}+\Gamma_{K^+K^-})\nonumber \\
\Gamma_0&=g_{\pi\pi}q_{\pi}(m_f)
\end{align}
for the $f_0(980)$ meson, with parameters from Ref. \cite{Ablikim:2004wn}.
If not mentioned otherwise, the parameters
are set to their PDG values~\cite{Tanabashi:2018oca}. The $\sigma$
meson contribution can be viewed as a phase space contribution to
the $\omega\pi\pi$ channel more than resonant contribution. Therefore,
the width is chosen to be large, see Tab.~\ref{tab:omegapipi}.

\begin{table}[b!]
\centering
\begin{tabular}{l|rr}
\toprule
Parameter & Fit value & PDG\\ \midrule
$m_{\omega''}$ & $1.69\pm 0.00919$ GeV & $1.670\pm 0.03$ GeV \\ 
$\Gamma_{\omega''}$ & $0.285\pm 0.0143$ GeV & $0.315\pm 0.035$ GeV \\ 
$m_{\sigma}$ & $0.6$ GeV & - \\ 
$\Gamma_{\sigma}$ & $1.0$ GeV & - \\ 
$g_{\omega''\omega\sigma}$ & 1. (fixed) & -  \\ 
$m_{f_0(980)}$ & 0.980 GeV & $0.990\pm 0.020$ GeV  \\ 
$\Gamma_{f_0(980)}$ & 0.1 GeV & 0.01-0.1 GeV  \\ 
$g_{\omega''\omega f_0(980)}$ & $0.883\pm 0.0616$ & - \\ 
$g_{\omega''}$ & $1.63\pm 0.0598$ & - \\ 
$\chi^2/\text{n.d.f}$ & \multicolumn{2}{c}{2.001} \\ \bottomrule
\end{tabular}
\caption{Fit values for the $\omega\pi\pi$ channel.}
\label{tab:omegapipi}
\end{table}

\subsubsection*{\boldmath$KK\pi$ (new)}

\begin{table}[t]
\centering
\begin{tabular}{l|cc|cc}
\toprule
fit value & $I$ & $i=1$ & $i=2$ & $i=3$ \\ \midrule
\multirow{2}{*}{$A_{I,i}$ in $\text{GeV}^{-1}$} & $I=0$ & 0 (fixed)  & $0.233\pm 0.020$ &
                                                             $0.0405\pm 0.0081$ \\ 
& $I=1$ & $-2.34\pm 0.15$ & $0.594\pm0.023$& $-0.018\pm 0.013$\\
\multirow{2}{*}{$\varphi_{I,i}$} & $I=0$ & $0$ (fixed) & $1.1\text{E-07} \pm 0.092$ &
  $5.19\pm 0.34$\\
& $I=1$ & $0$ (fixed)& $0.317\pm 0.056$ & $2.57\pm 0.32$ \\
\multirow{2}{*}{$m_{I,i}$ [GeV]}& $I=0$ & $1.019461$ (fixed)& $1.6334\pm
                                                      0.0065$ &
                                                                $1.957\pm 0.034$\\
& $I=1$ & 0.77526 (fixed)& 1.465 (fixed) & 1.720 (fixed) \\
\multirow{2}{*}{$\Gamma_{I,i}$ [GeV]}& $I=0$ & 0.004249 (fixed)&  $0.218\pm
                                                          0.013$&
                                                                  $0.267\pm
                                                                  0.032$\\
& $I=1$ & $0.1491$ (fixed) & $0.400$ (fixed) & $0.250$ (fixed)\\
\bottomrule
\end{tabular}
\caption{Fit values for the $KK\pi$ channel.}
\label{tab:KKpi}
\end{table}

Below 2~GeV center-of-mass energy the process \mbox{$e^+e^- \to KK\pi$} is
dominated by \linebreak\mbox{$e^+e^-\to KK^* \to K(K\pi)$} where $KK^*$ can be either
$K^0K^{*0}(890)$ or $K^\pm K^{*\mp}(890)$. We can relate the possible
final states through their isospin $I=0,1$ and can use the following
relations for the corresponding amplitudes
$A_{0,1}$~\cite{Davier:2010nc},
\begin{align}
K^+(K^-\pi^0)+K^-(K^+\pi^0)&: \quad \frac{1}{\sqrt{6}}(A_0-A_1),\nonumber \\
K^0_S (K^0_L\pi^0)+K^0_L(K^0_S\pi^0)&: \quad \frac{1}{\sqrt{6}}(A_0+A_1),\nonumber \\
K^0(K^-\pi^+)+\bar{K}^0(K^+\pi^-)&: \quad
                                 \frac{1}{\sqrt{3}}(A_0+A_1),\nonumber
  \\
K^+(\bar{K}^0\pi^-)+K^-(K^0\pi^+)&: \quad \frac{1}{\sqrt{3}}(A_0-A_1)~.
\end{align}
For the amplitudes with intermediate resonances, $e^+e^-\to V
\to KK^*$, we use the standard Breit-Wigner dsitribution
\begin{align}
A_{I}=\sum_i A_{I , i}
\frac{m_{I,i}^2e^{\varphi_{I,i}}}{m_{I,i}^2-\hat{s}-i\sqrt{\hat{s}}\Gamma_{I,i}}~.
\end{align}
In the energy range we are dealing with, we expect the resonances to
be $\phi (1680)$ and $\phi(2170)$ for $I=0$ and $\rho(1450)$ and
$\rho(1700)$ for $I=1$. The lower resonances $\rho(770)$ and
$\phi(1020)$ are not considered in the energy range of the fit and we
set their couplings to zero. Furthermore, we fix the mass and the
width of the intermediate $K^*$ resonance to $m_{K^*}=0.8956$~GeV and
$\Gamma_{K^*}=0.047$~GeV and use a $p$-wave Breit-Wigner
propagator of the form
\begin{align}
\text{BW}_{K^*}(s)=\frac{g_{K^*K\pi} m_{K^*}^2}{m_{K^*}^2-s-i\sqrt{s}\Gamma(s)},
\end{align}
with the $s$-dependent width 
\begin{align}
\Gamma(s)=\Gamma_{K^*}\frac{\sqrt{s}}{m_{K^*}}\left(\frac{\beta(s,m_1,m_2)^2}{\beta(m_{K^*},m_1,m_2)^2}\right)^{3/2}.
\end{align}
where $m_1,m_2$ are the decay products of the $K^*$ state and 
\begin{align}
\beta(s,m_1,m_2)= \left(1-\frac{(m_1+m_2)^2}{s}\right)^{1/2}\left(1-\frac{(m_1-m_2)^2}{s}\right)^{1/2}
\end{align}
is their velocity in the rest frame of $K^*$. The $K^*K\pi$ coupling is
given by
\begin{align}
g_{K^*K\pi}=\sqrt{6\pi m_{K^*}^2/(0.5 m_{K^*}\beta(m_{K^*}^2,m_{K^\pm},m_{\pi^\pm}))^3\Gamma_{K^*}}=5.37392360229~.
\end{align}
Furthermore, we include a small $\phi\pi^0$ contribution for final
states including neutral pions by adding the $\phi\pi^0$ cross section
obtained by the $\phi\pi$ fit and the corresponding branching
fractions $\text{BR}(\phi(1020)\to K^0_LK^0_S)=0.342$ and
$\text{BR}(\phi(1020)\to K^+K^-)=0.489$.  We perform a simultaneous
fit to all possible final states in order to obtain the fit parameters
of the amplitudes $A_{0,1}$. The fit values can be found in
Tab.~\ref{tab:KKpi}.

We show all numerical best-fit solutions as blue lines for all final
states in Figs.~\ref{fig:fit1}, \ref{fig:fit2}, and \ref{fig:fit3}. The
error bars on the data are dominated by statistical uncertainties. All
fits describe the most recent data sets over the entire range
shown.

\subsubsection*{Error bands}

\begin{figure}
\begin{subfigure}[b]{0.5\linewidth}
\begin{center}
\includegraphics[width=0.95\linewidth]{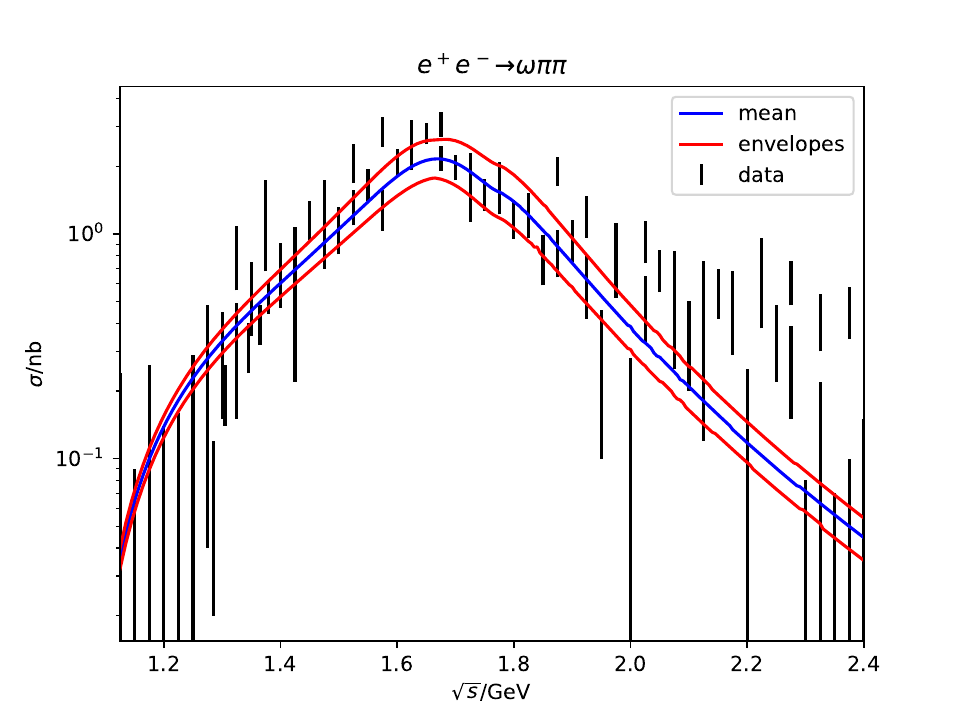}
\end{center}
\end{subfigure}
\begin{subfigure}[b]{0.5\linewidth}
\begin{center}
\includegraphics[width=0.95\linewidth]{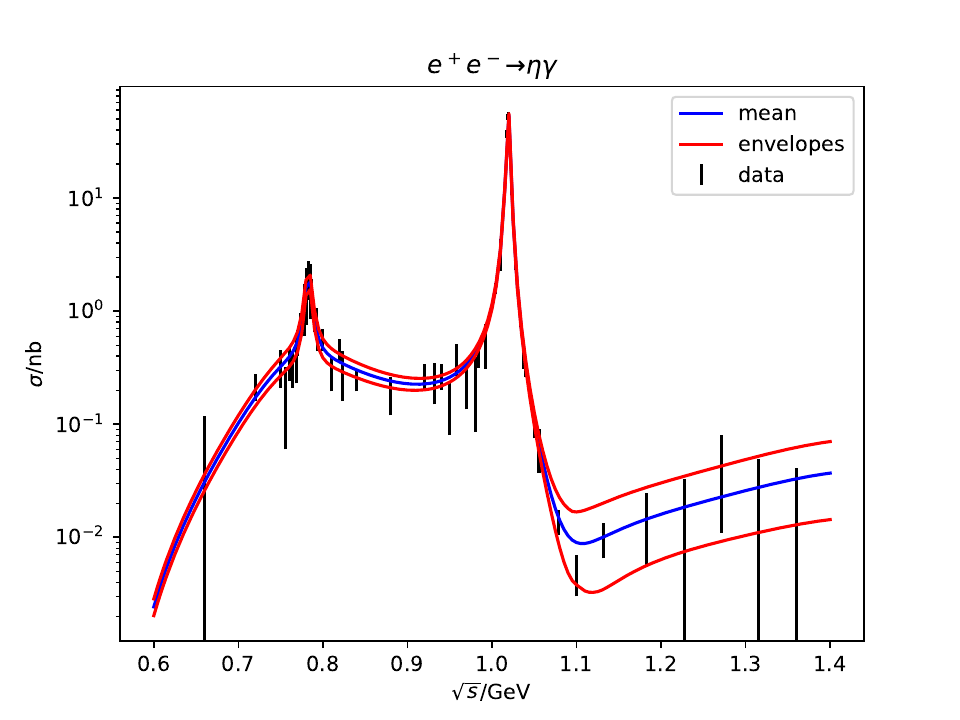}
\end{center}
\end{subfigure}
\begin{subfigure}[b]{0.5\linewidth}
\begin{center}
\includegraphics[width=0.95\linewidth]{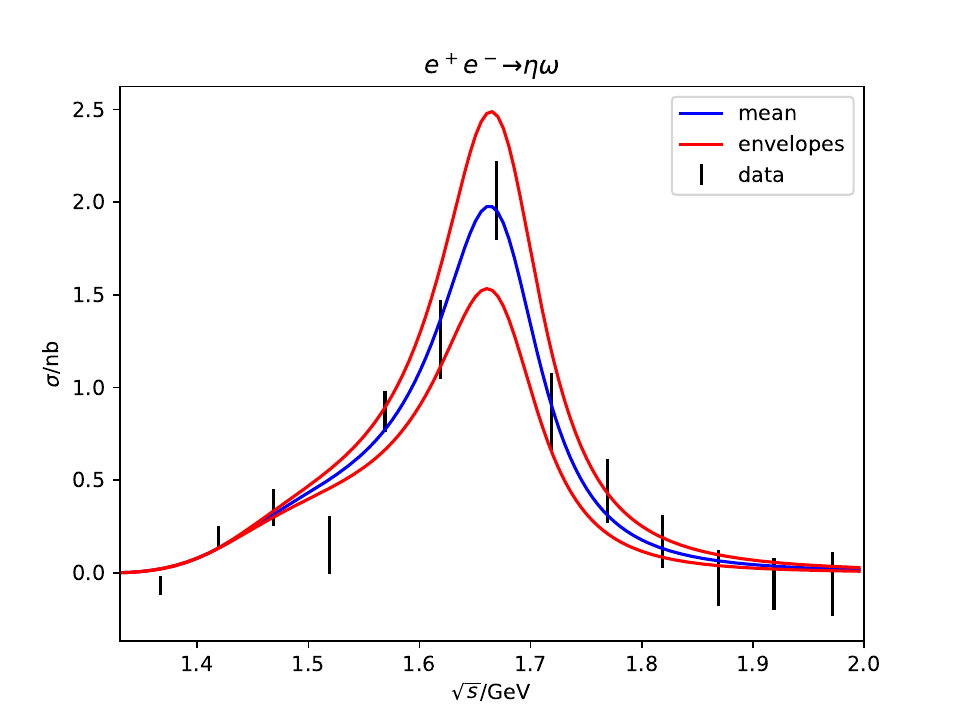}
\end{center}
\end{subfigure}
\begin{subfigure}[b]{0.5\linewidth}
\begin{center}
\includegraphics[width=0.95\linewidth]{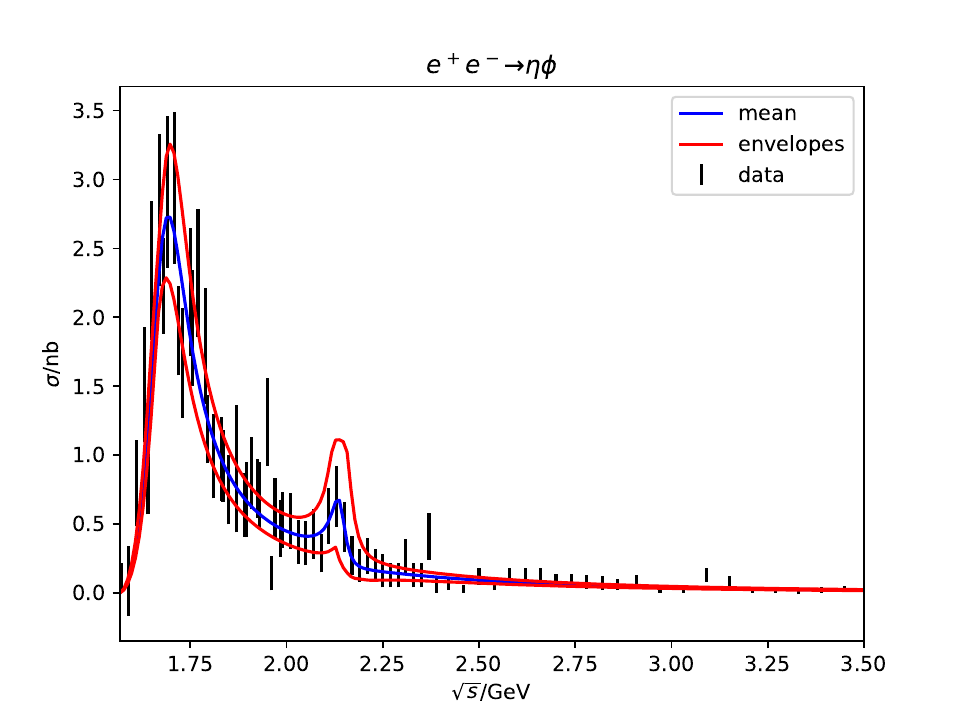}
\end{center}
\end{subfigure}
\begin{subfigure}[b]{0.5\linewidth}
\begin{center}
\includegraphics[width=0.95\linewidth]{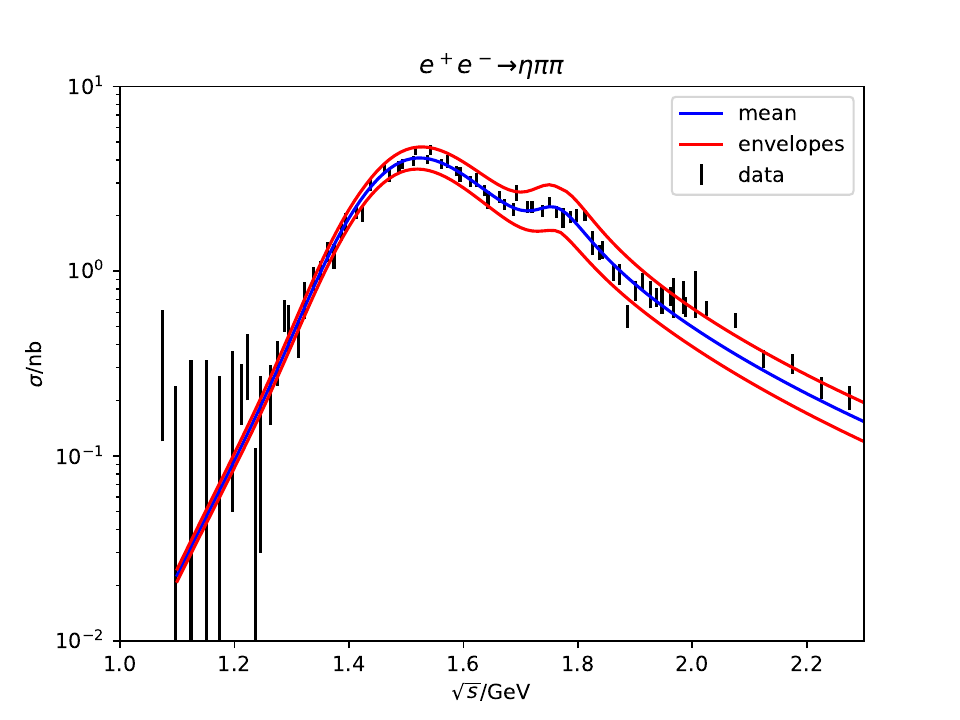}
\end{center}
\end{subfigure}
\begin{subfigure}[b]{0.5\linewidth}
\begin{center}
\includegraphics[width=0.95\linewidth]{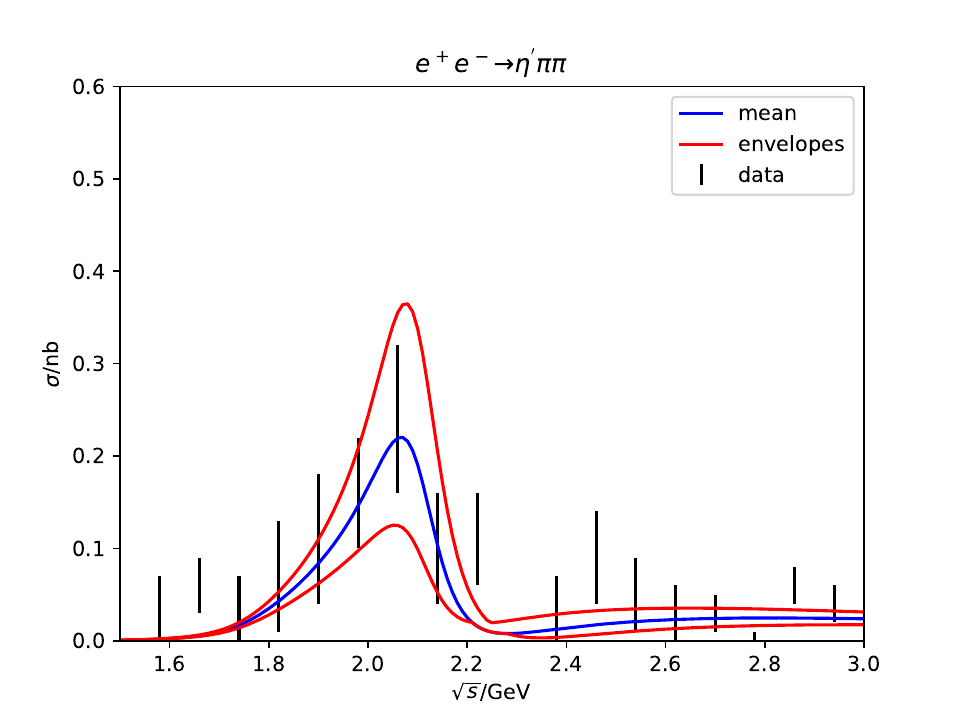}
\end{center}
\end{subfigure}
\begin{subfigure}[b]{0.5\linewidth}
\begin{center}
\includegraphics[width=0.95\linewidth]{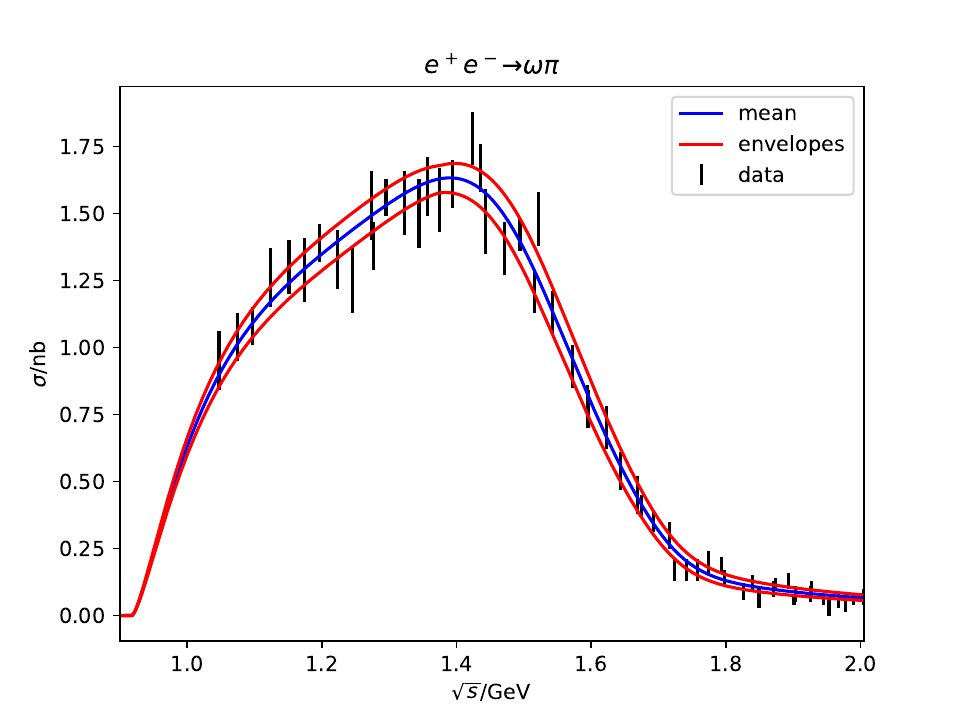}
\end{center}
\end{subfigure}
\begin{subfigure}[b]{0.5\linewidth}
\begin{center}
\includegraphics[width=0.95\linewidth]{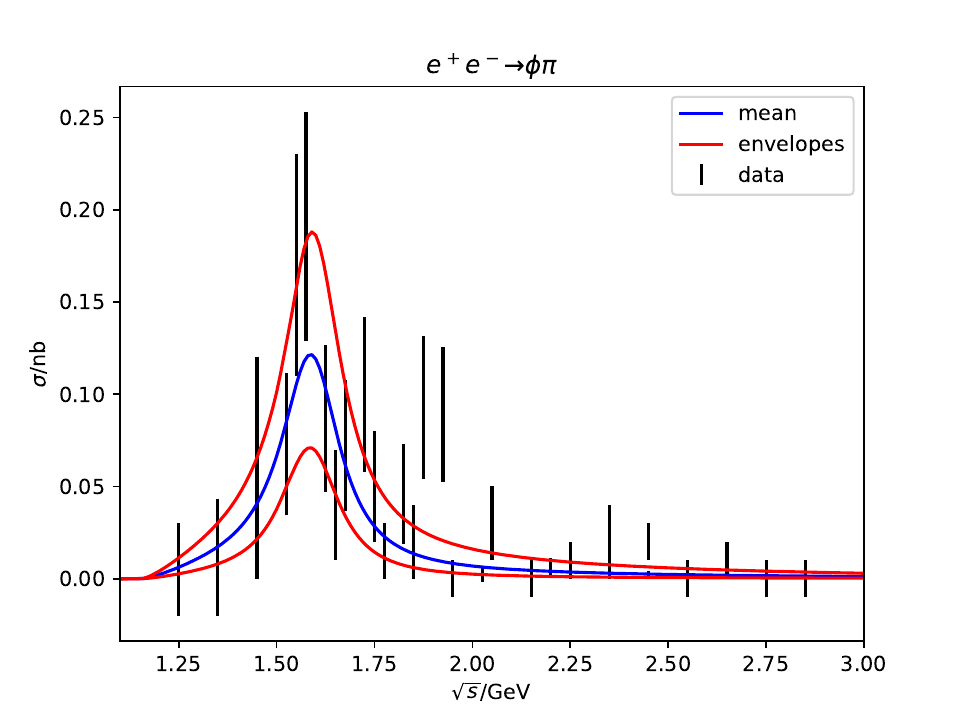}
\end{center}
\end{subfigure}
\caption{Cross sections for hadronic final states with error envelopes.}
\label{fig:fit1}
\end{figure}

\begin{figure}
\begin{subfigure}[b]{0.5\linewidth}
\begin{center}
\includegraphics[width=0.95\linewidth]{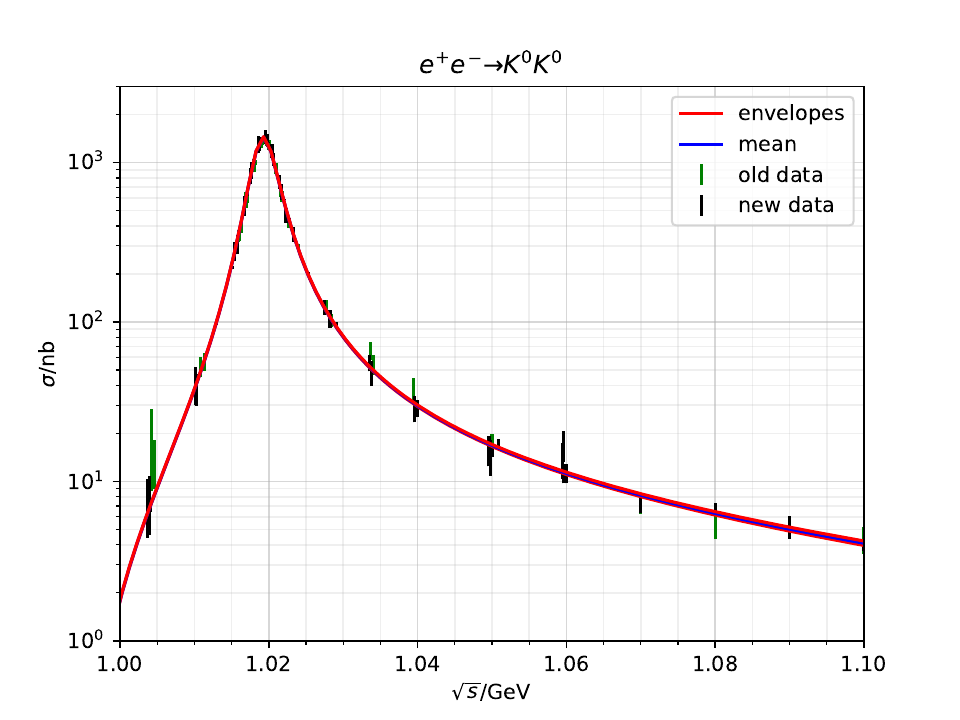}
\end{center}
\end{subfigure}
\begin{subfigure}[b]{0.5\linewidth}
\begin{center}
\includegraphics[width=0.95\linewidth]{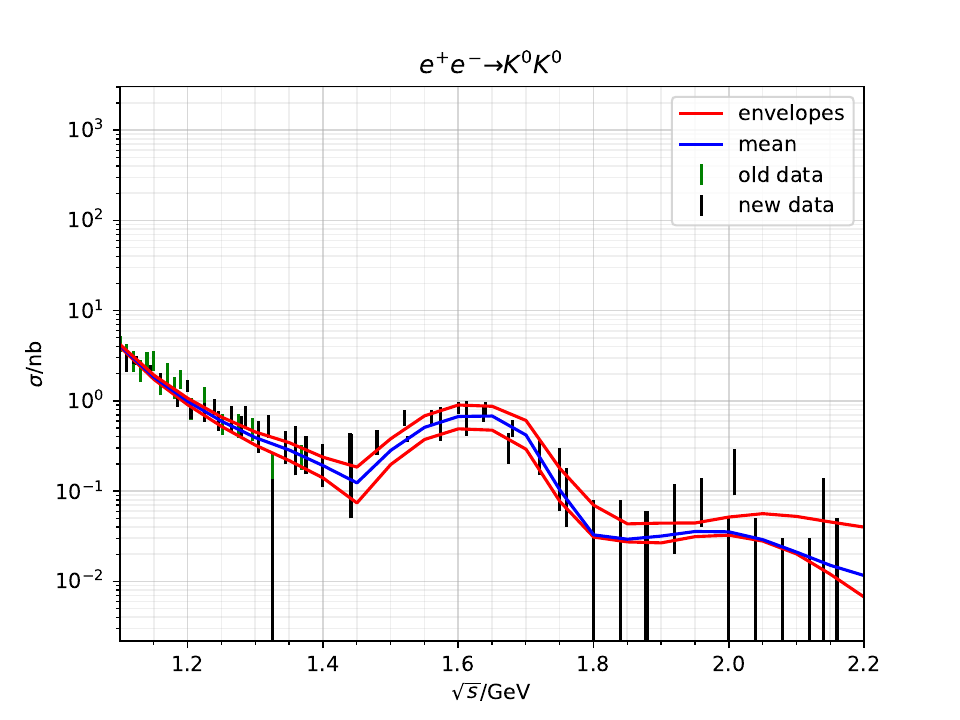}
\end{center}
\end{subfigure}
\begin{subfigure}[b]{0.5\linewidth}
\begin{center}
\includegraphics[width=0.95\linewidth]{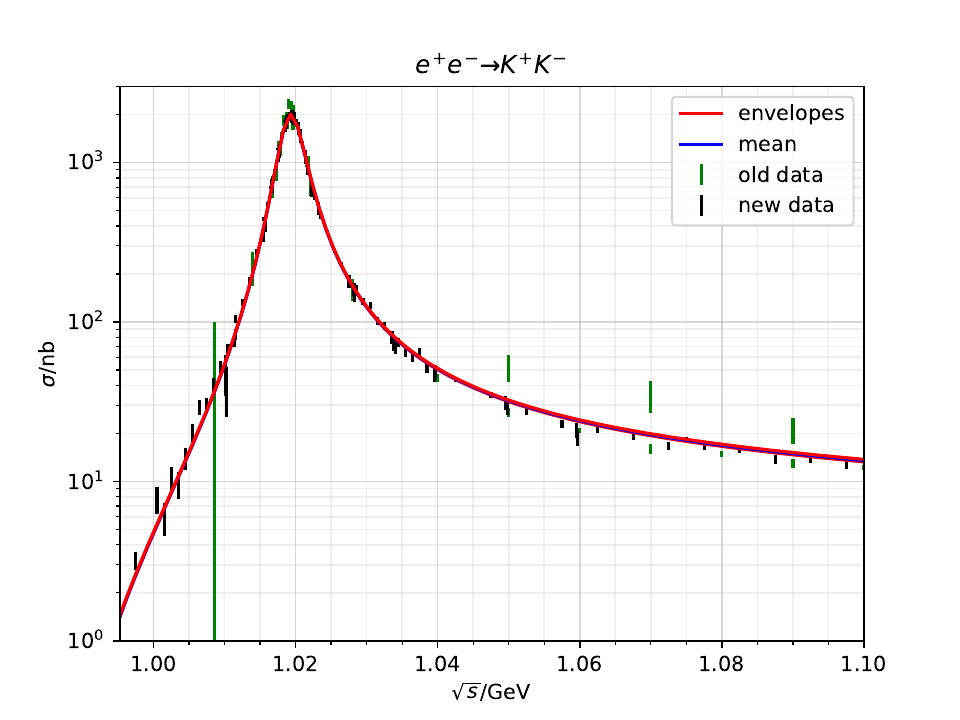}
\end{center}
\end{subfigure}
\begin{subfigure}[b]{0.5\linewidth}
\begin{center}
\includegraphics[width=0.95\linewidth]{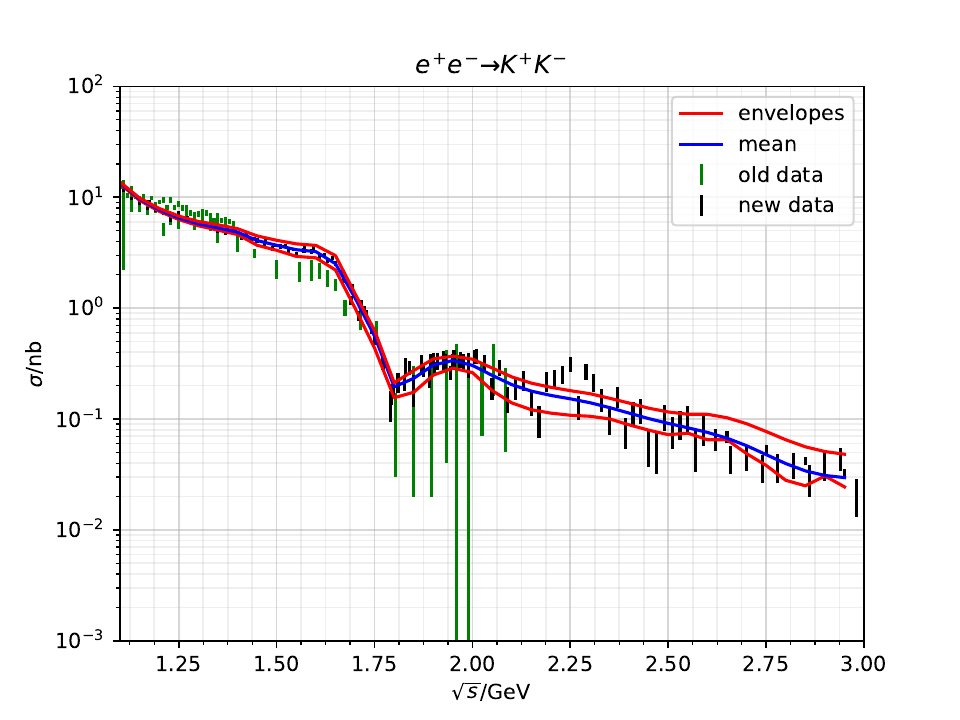}
\end{center}
\end{subfigure}
\begin{subfigure}[b]{0.5\linewidth}
\begin{center}
\includegraphics[width=0.95\linewidth]{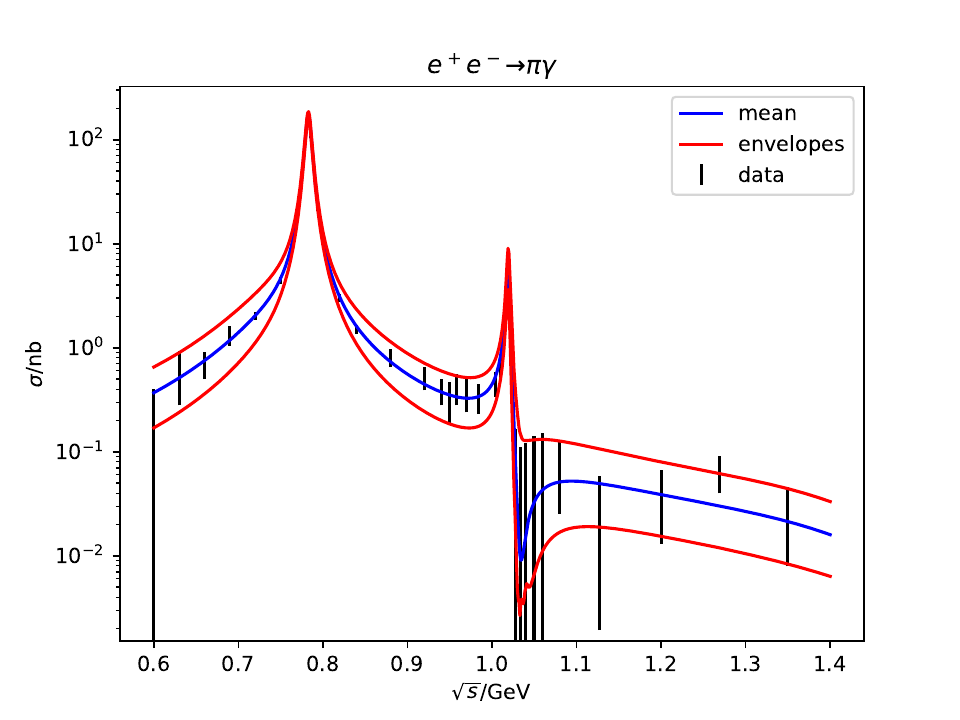}
\end{center}
\end{subfigure}
\begin{subfigure}[b]{0.5\linewidth}
\begin{center}
\includegraphics[width=0.95\linewidth]{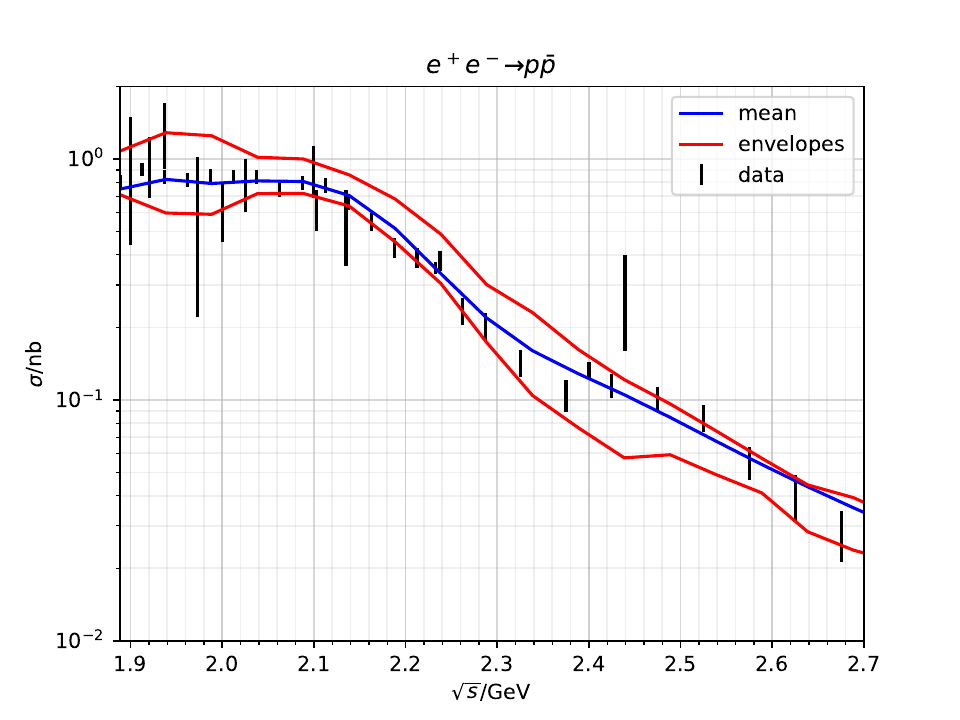}
\end{center}
\end{subfigure}
\begin{subfigure}[b]{0.5\linewidth}
\begin{center}
\includegraphics[width=0.95\linewidth]{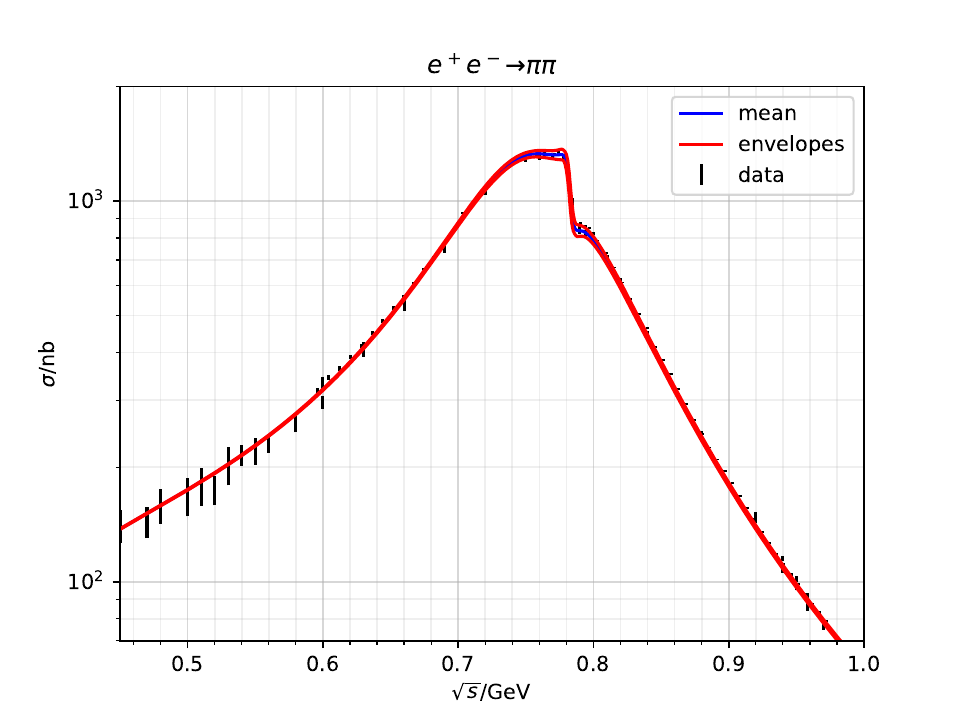}
\end{center}
\end{subfigure}
\begin{subfigure}[b]{0.5\linewidth}
\begin{center}
\includegraphics[width=0.95\linewidth]{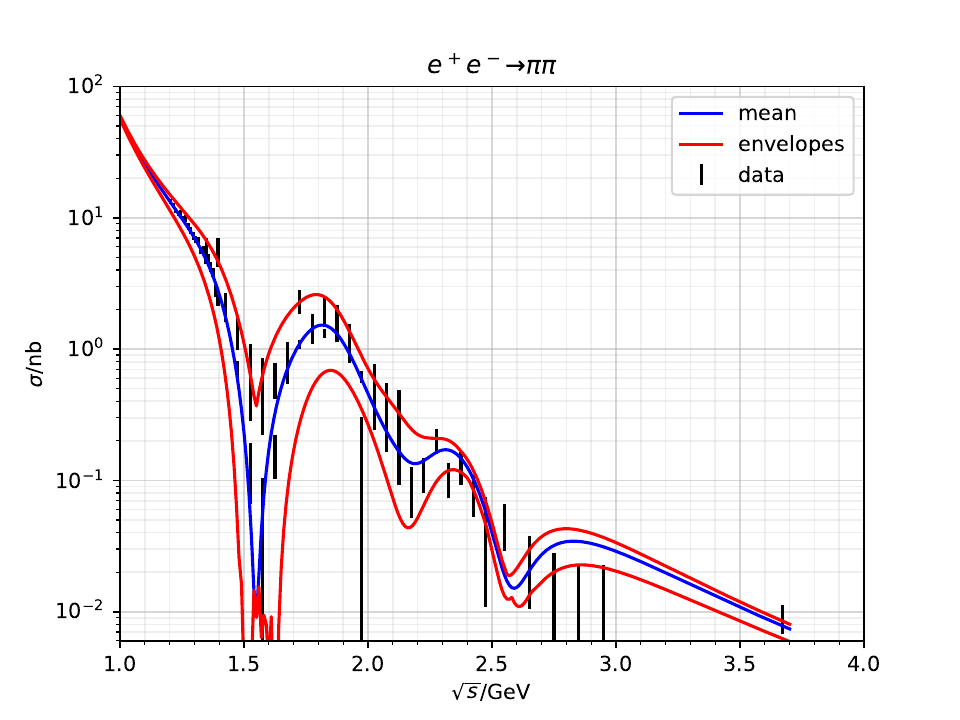}
\end{center}
\end{subfigure}
\caption{Cross sections for hadronic final states with error envelopes.}
\label{fig:fit2}
\end{figure}

\begin{figure}
\begin{subfigure}[b]{0.5\linewidth}
\begin{center}
\includegraphics[width=0.95\linewidth]{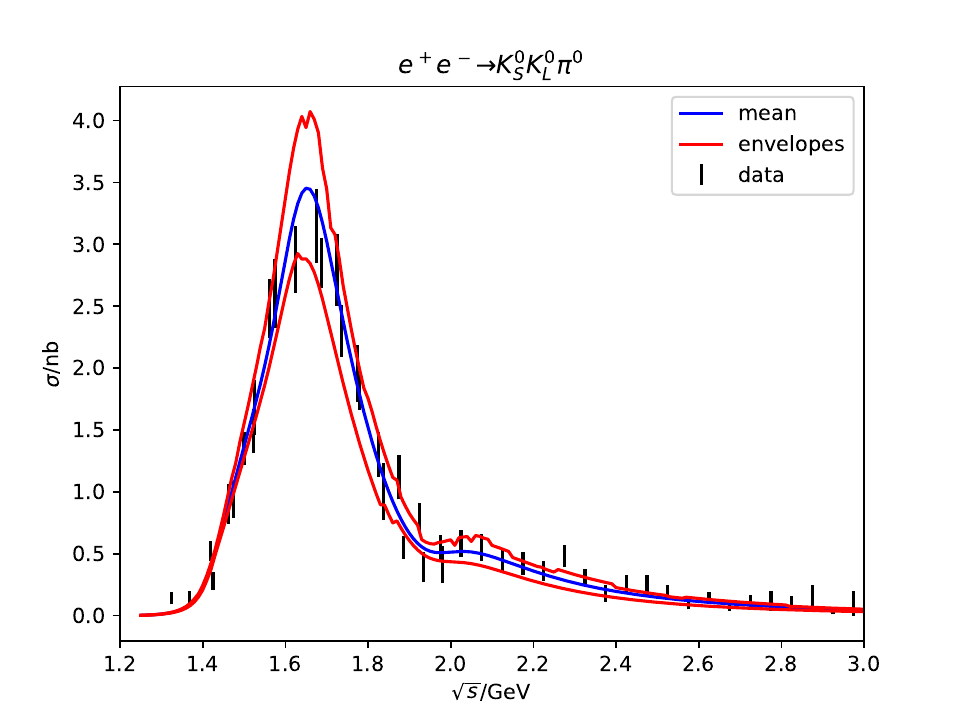}
\end{center}
\end{subfigure}
\begin{subfigure}[b]{0.5\linewidth}
\begin{center}
\includegraphics[width=0.95\linewidth]{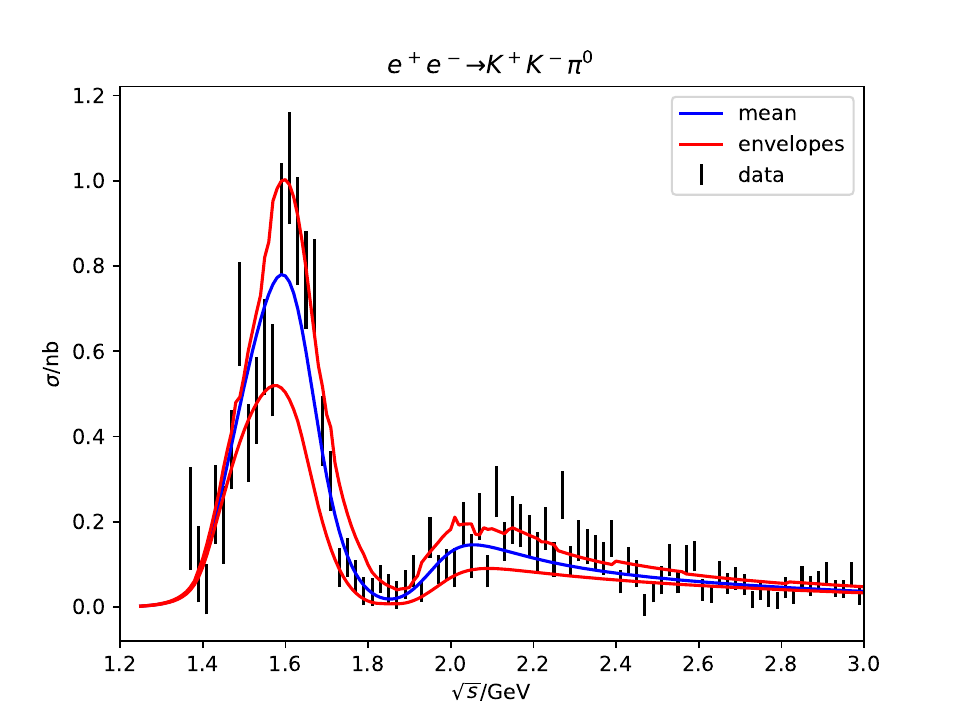}
\end{center}
\end{subfigure}
\begin{subfigure}[b]{0.5\linewidth}
\begin{center}
\includegraphics[width=0.95\linewidth]{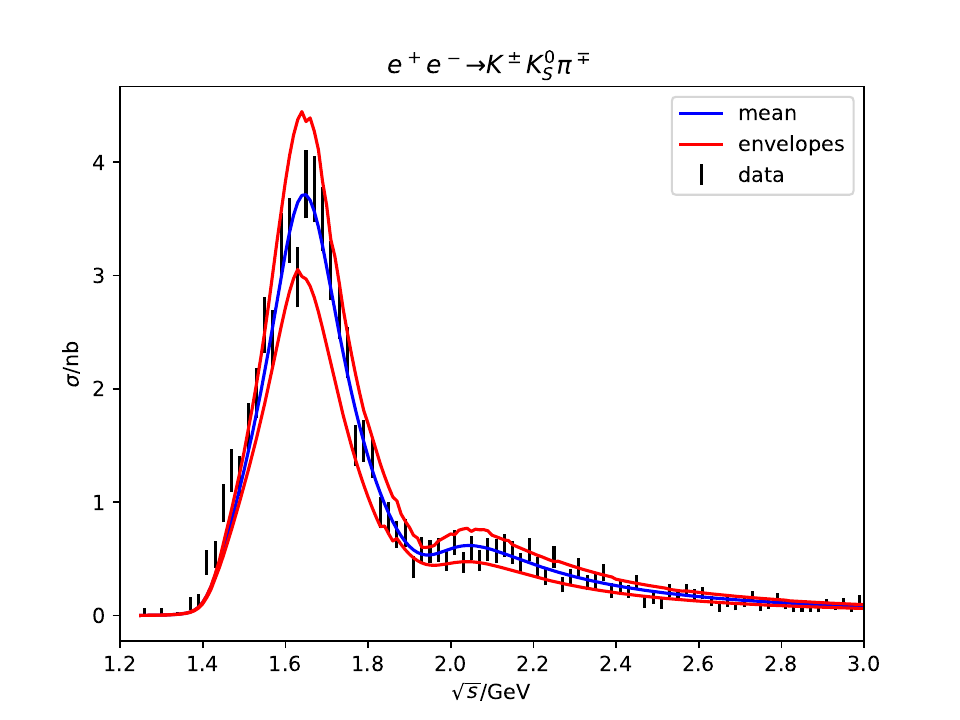}
\end{center}
\end{subfigure}
\begin{subfigure}[b]{0.5\linewidth}
\begin{center}
\includegraphics[width=0.95\linewidth]{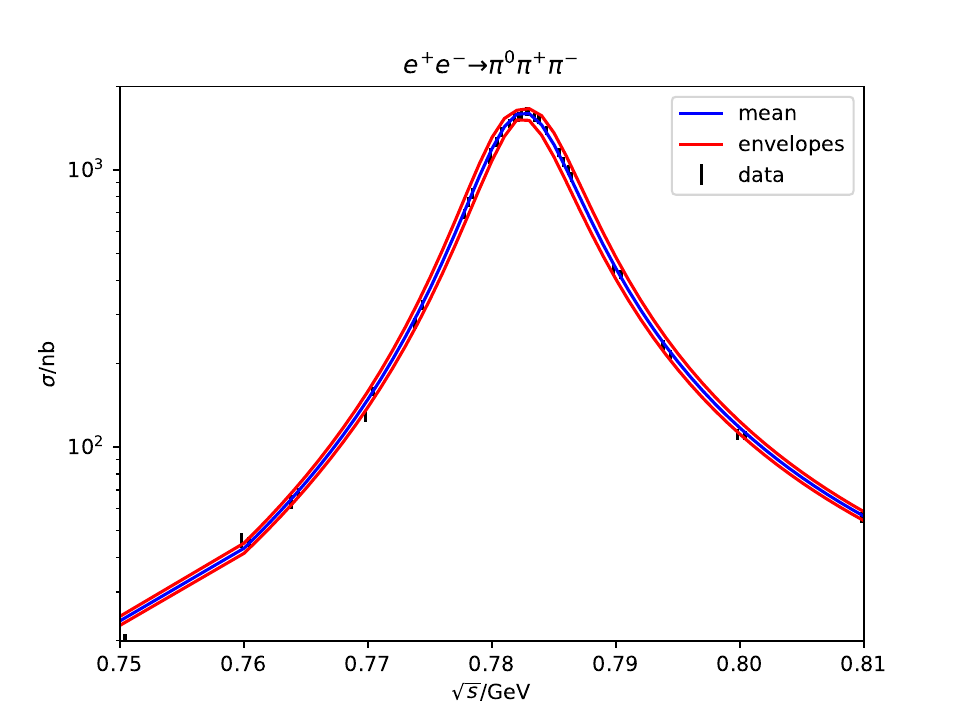}
\end{center}
\end{subfigure}
\begin{subfigure}[b]{0.5\linewidth}
\begin{center}
\includegraphics[width=0.95\linewidth]{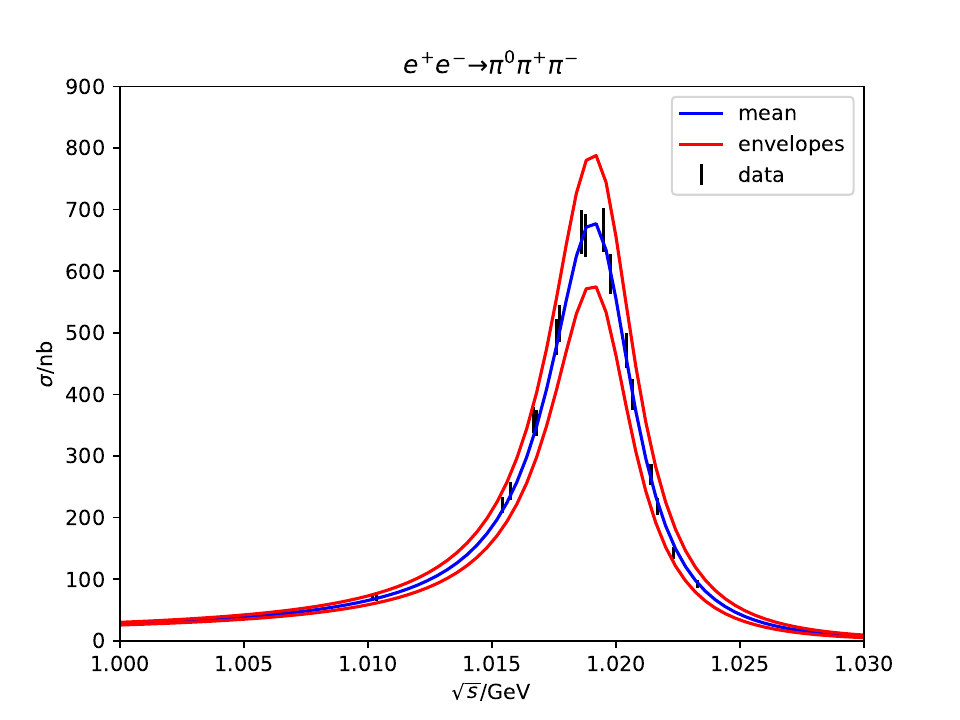}
\end{center}
\end{subfigure}
\begin{subfigure}[b]{0.5\linewidth}
\begin{center}
\includegraphics[width=0.95\linewidth]{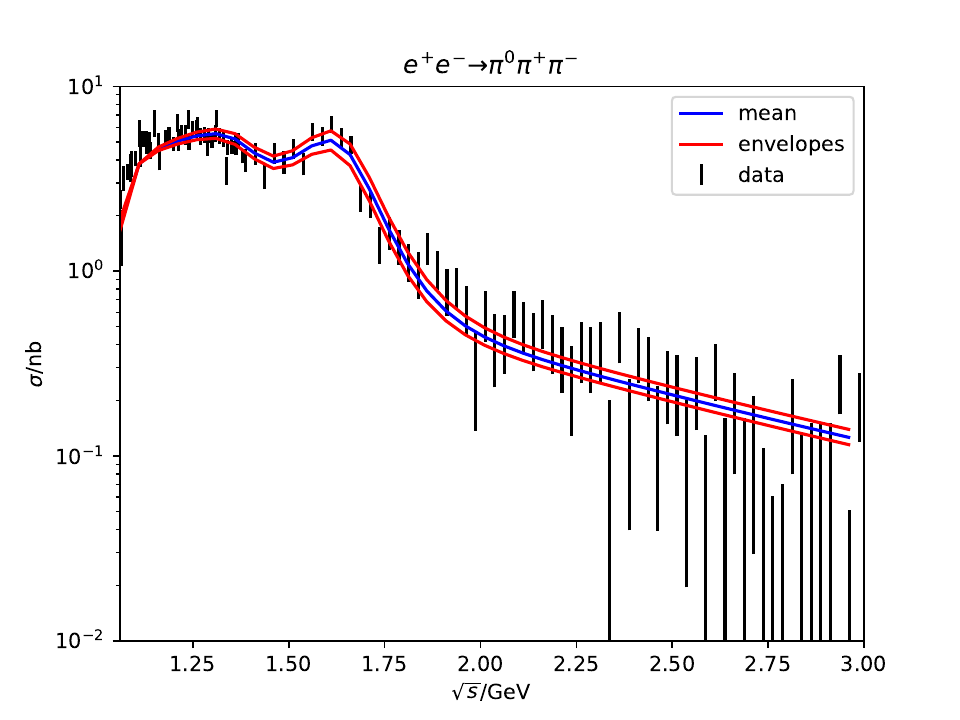}
\end{center}
\end{subfigure}
\begin{subfigure}[b]{0.5\linewidth}
\begin{center}
\includegraphics[width=0.95\linewidth]{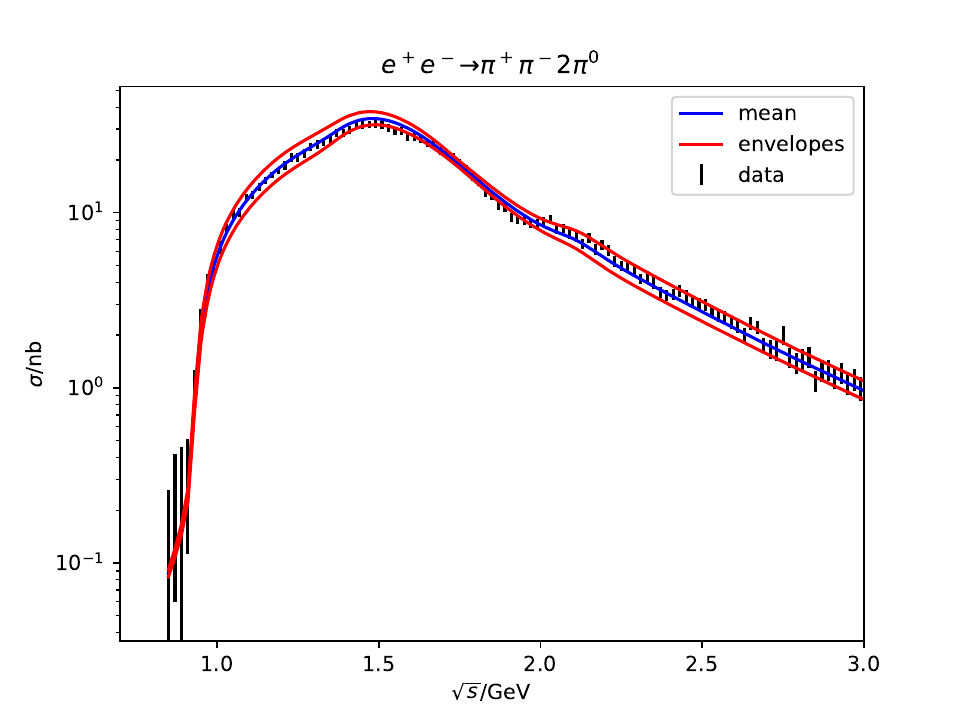}
\end{center}
\end{subfigure}
\begin{subfigure}[b]{0.5\linewidth}
\begin{center}
\includegraphics[width=0.95\linewidth]{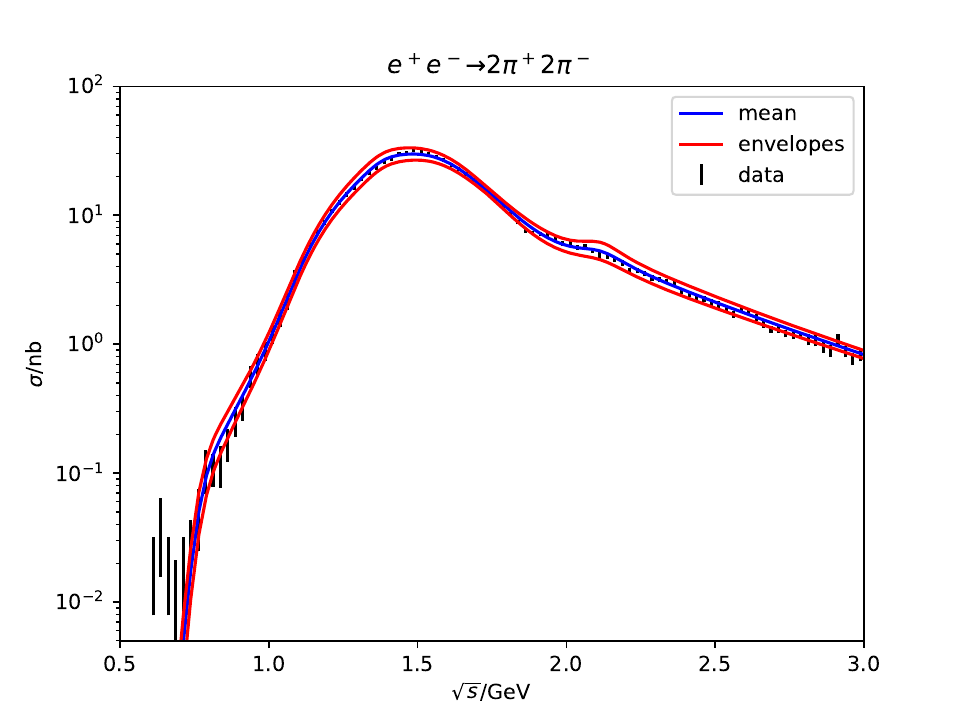}
\end{center}
\end{subfigure}
\caption{cross sections for hadronic final states with error envelopes.}
\label{fig:fit3}
\end{figure}

In addition to the central values of the relevant parameters
describing the $e^+e^-$ data we also estimate the error bands for the
relevant processes. The reason is that some of the channels are rather
poorly measured, and it is important to propagate these uncertainties
through the analysis. Because most fit parameters are physical
parameters appearing in the analytic description of the $e^+ e^-$
cross sections, such as masses or widths or rates, we do not find them
suitable for a proper statistical analysis. For instance a total cross
section measurement will lead to uncontrolled correlations between
widely different phase space regions in the fit, where the different
phase space regions are crucial to describe the dark matter spectra
for a variable dark matter mass. Examples for the impact of a known
form of the energy dependence of the scattering process on poorly
measured phase space regions are the $\eta \pi\pi$ channel in
Fig.~\ref{fig:fit1}, the $\pi\pi$ channel in Fig.~\ref{fig:fit2}, or
the $3\pi$ channel in Fig.~\ref{fig:fit3}.

Instead, we define envelopes by varying a sub-set of fit parameters
around their mean value within their uncertainty provided our python
\textsc{IMinuit}~\cite{iminuit,1975CoPhC..10..343J} fit or as stated in papers.  For poorly resolved peak
structures as in the $\eta'\pi\pi$, $\phi\pi$, and $\eta\omega$ case
or higher resonances as in $\eta\phi$ and $KK\pi$, we do not vary any
widths and only some masses, since they are determined from the peak
structure and bias the off-peak spectrum through correlations. The
contribution of phases to our envelopes is only considered if no other
set of parameters is sufficient to describe the measurement
uncertainties.  For channels with simple parametrizations with fixed
masses and widths and floating peak cross sections and phases as in
the case of $\pi\gamma$~\cite{Achasov:2016bfr} and
$\eta\gamma$~\cite{Achasov:2006dv}, we vary all peak cross sections
and the phases of the $\phi$ and $\omega$ resonance, respectively. In
these cases, we see that away from the resonance region the error
envelopes increase.  For precisely measured phase space regions, we
consider the full set of parameters describing these regions.  These
are usually large peak structures such as the $\phi \to KK$ and $\rho \to
\pi\pi$ resonances in Fig.~\ref{fig:fit2} or the $\omega,\rho \to
3\pi$ peak around 0.78~GeV in Fig~\ref{fig:fit3}. Those resolved
regions turn out to be well described and are stable against
variations of the parameters, so they give only small envelopes.

It can be challenging or nearly impossible to obtain consistent
envelopes for some channels, where one parametrization is used for
several sub-channels as in the case of $KK$ and
$p\bar{p}/n\bar{n}$. As long as the shape of the data is the same as
in the case of $4\pi$, $KK\pi$ and the $\phi$ resonance region in the
$KK$ channel, this does not cause any problems. Here we can assume
that a parameter and its variation influence the fit curve in the same
way. However, for energies above 1.4~GeV in the $KK$ channel, the
trend of the data of $K^+K^-$ and $K^0\bar{K}^0$ is completely
different. Therefore, already the fit to the data is challenging and
only possible by allowing for more resonance fit parameters in the
parametrization~\cite{Czyz:2010hj}.  A variation of the parameter
might influence both channels differently and it is not clear that an
extremal value in the one case is also extremal in the other. This
tension of both data sets causes too small error bands for energies
above 1.8~GeV. For the $p\bar{p}/n\bar{n}$ channel, we do not have
sufficient data for $n\bar{n}$ to describe this channel properly as
already described in Ref.~\cite{Czyz:2014sha}.

\clearpage

\bibliography{literature}

\end{document}